\documentclass[aps,prd,showpacs,nofootinbib,superscriptaddress,preprint,eqsecnum,tightenlines]{revtex4}

\usepackage{graphicx}
\usepackage{amsmath}
\usepackage{amsfonts}
\usepackage{amssymb}
\usepackage{amsthm}
\usepackage{newlfont}
\usepackage{dsfont}
\usepackage{newlfont}
\usepackage{hyperref}
\usepackage{color}
\usepackage{natbib} 

\def\bea{\begin{eqnarray}}
\def\eea{\end{eqnarray}}
\def\be{\begin{equation}}
\def\ee{\end{equation}}

\def\ba{\nopagebreak[3]\begin{eqnarray}}
\def\ea{\end{eqnarray}}

\def\d{{\rm d}}

\newcommand{\teta}{\rlap{\lower2ex\hbox{$\,\tilde{}$}}\eta{}}

\usepackage{enumerate}

\def\be{\begin{equation}}
\def\ee{\end{equation}}

\def\d{\mathrm{d}}

\begin{document}

\title{Energy in first order 2+1 gravity}
\author{Alejandro Corichi}\email{corichi@matmor.unam.mx}
\affiliation{Centro de Ciencias Matem\'aticas, Universidad Nacional Aut\'onoma de
M\'exico, UNAM-Campus Morelia, A. Postal 61-3, Morelia, Michoac\'an 58090,
Mexico.}
\affiliation{Center for Fundamental Theory, Institute for Gravitation and the Cosmos,
Pennsylvania State University, University Park
PA 16802, USA.}
\author{Ira\'\i s Rubalcava-Garc\'\i a}
\email{irais11@gmail.com}
\affiliation{Facultad de Ciencias F\'{\i}sico Matem\'aticas, Universidad Aut\'onoma de Puebla, A. Postal 1152, 72001 Puebla, Puebla, Mexico.}
\affiliation{Instituto de F\'{\i}sica y
Matem\'aticas,  Universidad Michoacana de San Nicol\'as de
Hidalgo, Morelia, Michoac\'an, Mexico.}
\affiliation{Centro de Ciencias Matem\'aticas, Universidad Nacional Aut\'onoma de
M\'exico, UNAM-Campus Morelia, A. Postal 61-3, Morelia, Michoac\'an 58090,
Mexico.}

\begin{abstract}
We consider $\Lambda$=0 three dimensional gravity with asymptotically flat boundary conditions. This system was studied  
by Ashtekar and Varadarajan within the second order formalism --with metric variables--  who showed that the Regge-Teitelboim formalism yields a consistent Hamiltonian description where, surprisingly, the energy is bounded from below {\it and} from above. The energy of the spacetime is, however, determined up to an arbitrary constant. The natural choice was to fix that freedom such that Minkowski spacetime has zero energy. More recently, Marolf and Pati\~no 
started from the Einstein-Hilbert action supplemented with the Gibbons-Hawking term and showed that, in the 2+1 decomposition of the theory, the energy is shifted from the Ashtekar-Varadarajan analysis in such a way that Minkowski spacetime possesses  a {\it negative} energy. In this contribution we consider the first order formalism, where the fundamental variables are a $so(2,1)$ connection $w_a{^{I}}_{J}$ and a triad $e_a^I$. We consider two actions. A natural extension to 3 dimensions of the consistent action in $4D$ Palatini gravity is shown to be finite and differentiable. For this action, the 2+1 decomposition (that we perform using two methods) yields a Hamiltonian boundary term that corresponds to energy. It assigns  zero energy to Minkowski spacetime. We then put forward a totally gauge invariant action, and show that it is also well defined and differentiable. Interestingly, it turns out to be related, on shell, to the 3D Palatini action by an additive constant in such a way that its associated energy is given by the Marolf-Pati\~no expression. Thus, we conclude that, from the perspective of the first order formalism, Minkowski spacetime can consistently have {\it either}, zero, {\it or} a negative energy equal to $-1/4G$, depending on the choice of consistent action employed as starting point.
\end{abstract}

\pacs{04.20.-q, 04.20.Fy, 04.20.Ha}
\maketitle


\section{Introduction}
\label{sec:1}



Idealized and reduced models have been useful in analyzing and studying, in a simplified arena, some aspects of (3+1) general relativity. To be more precise, one can consider the sector of Einstein theory that is invariant under certain symmetries, that sometimes becomes tractable, in order to gain some insight into the full theory. An outstanding example of such simplified model is the (2+1)-dimensional case which, apart from being much simpler than the (3+1) case, it has been `solved' in many different contexts and by different approaches \cite{deser,Witten1988}. It is then natural to explore and compare the resulting formalism with the hope of learning something new about the full (3+1) case. 


The issue that we shall here consider is the definition of gravitational energy.
This endeavor is certainly not new and a sizeable amount of literature has been
devoted to this topic in both 4D \cite{energy} and 3D gravity \cite{jackiw}. In the case of
3D gravity, the present situation is not devoid of some tension. More precisely,
the first systematic study of asymptotically flat ($\Lambda=0$) boundary 
conditions was first put forward by Ashtekar and Varadarajan in
\cite{Ashtekar-Varadarajan1994}. They made precise the notion of asymptotically flat boundary 
conditions for the canonical theory and concluded, within the Regge-Teitelboim 
formalism \cite{regge-teitel} that the canonical energy is not only bounded from 
below, as one could have expected, but it is also bounded from above. This 
unexpected feature has some interesting consequences when considering the
quantum theory \cite{aa:quantum}. The Regge-Teitelboim formalism suffers,
nevertheless, from an ambiguity in the definition of the value of the energy it
assigns to, say, its lowest energy configuration. The ambiguity comes from the
fact that one could add an arbitrary constant to the Hamiltonian and the
formalism is still fully consistent. In the case of 3+1 gravity, 
this special configuration is precisely Minkowski spacetime and it is customary
to assign to it a {\it zero} value of energy. This choice is fully justified and
is not subject to any controversy. The same is not true for the 3D case. In
\cite{Ashtekar-Varadarajan1994} the authors chose the same convention and assigned
zero energy to 2+1 Minkowski spacetime. 

In \cite{Marolf-Patino2006} Marolf and Pati\~no followed a different approach. 
They started from a well defined second order action for the gravitational field
consisting of the standard Einstein-Hilbert action plus a boundary term given 
by the Gibbons-Hawking term. After a 2+1 decomposition they obtained the boundary
contribution to the Hamiltonian and found that there is an extra term that
``shifts'' the value of the energy in such a way that Minkowski spacetime is
assigned a negative value equal to $-1/4G$, and the upper bound on energy is zero.
Even when this result might appear to be counter-intuitive from the perspective of 3+1 gravity, there are several argument to support this behaviour. First, one should note that the gravitational constant $G$ in three spacetime dimensions has dimensions of inverse mass, so in this case one {\it does} have a mass scale even for vacuum gravity. Second, the asymptotic conditions at infinity are such that
there is a preferred notion of time translation. The symmetry group is much more restricted, and it is not strange to assign a non-zero ADM momentum to this preferred frame \cite{deser2,symmetry}.\footnote{In 3+1 gravity a non-zero value for the ADM four-momentum would select a preferred frame thus violating asymptotic Lorentz invariance.} Finally, the asymptotic spatial geometry corresponding to configurations where the energy approaches its limiting (upper) value correspond to two dimensional conical defects that ``close up". It is not then unnatural to assign a zero energy to such spatially closed spacetimes
\cite{Marolf-Patino2006}.

The use of first order variables for gravity in 3+1 dimensions has proven to be rather convenient. Apart from the necessity to consider them when coupling Fermions, they allow for a simple well defined action \cite{aes} in the case of
asymptotically flat configurations. A natural question is whether a corresponding
action principle can be defined for $2+1$ gravity. Here the main variables 
would be a co-triad $e_a^I$, together with a connection ${w_{aI}}^J$ taking
values in the Lie algebra of $SO$(2,1). 

The purpose of this manuscript is to
address several of these issues. First we extend the results of \cite{aes}
to three dimensions and derive the asymptotically flat conditions for the 
first order variables. Then, we prove that the 3-dimensional Palatini action with boundary term, which 
give us the same equations of motion that the 3-dimensional Einstein-Hilbert action, has a well posed 
action principle. That is, it is finite and differentiable under the asymptotically flat boundary 
conditions. Moreover, we define a new action principle by introducing an additional boundary term to the 
action. This new action is explicitly Lorentz invariant and, as we prove in detail, it is equivalent to the Einstein-Hilbert action with a Gibbons-Hawking term 
of \cite{Marolf-Patino2006}. The next step is to consider the covariant 
Hamiltonian formulation (CHF) defined by these two action principles and explore 
some of its relevant quantities. 
In particular, we prove that the energy is bounded from 
below and above,  for asymptotically 3-dimensional flat space-times, in agreement 
with previous results in the metric 
variables via Regge-Teitelboim methods \cite{Ashtekar-Varadarajan1994}. Although 
the CHF provides an elegant and short 
derivation for the energy (and other relevant symmetry generators as 
discussed in \cite{crv1,crv2}), the energy is only determined up to a constant, 
that shifts the region in which the energy is defined. 

Next, we consider the 2+1 decomposition of the two first order actions. 
We follow two different strategies. The first one, that we shall call the ``Witten" approach (See 
\cite{Witten1988} and \cite{romano} for details), exploits the fact that the bulk action has the 
structure of a $BF$ theory, where no underlying spacetime metric is assumed. The second approach,
as put forward by Barbero and Varadarajan \cite{Barbero-Varadarajan1994}, uses the fact that there is an 
underlying metric structure, and resembles the 3+1 first order case (as described in \cite{romano}). In 
both cases, we show that the resulting canonical theories are well defined and obtain the Hamiltonian from 
the corresponding boundary terms. We find that the energy
associated to the spacetime depends on the choice of action principle, differing by a constant. For the 
simplest Palatini action, the interval in which the energy is defined is positive,
and assigns a zero value of energy to Minkowski spacetime. For the fully gauge invariant action, 
we shall show that the energy is always negative and coincides with the values assigned by Marolf 
and Pati\~no. Thus,  from the perspective of the first order formalism, Minkowski spacetime can 
consistently have {\it either}, zero, 
{\it or} a negative energy equal to $-1/4G$, depending on the choice of consistent action employed as a 
starting point.



The structure of the manuscript is as follows. In Sec.~\ref{sec:2} we introduce the notion a asymptotic 
flatness for the first order variables. In Sec.~\ref{sec:3} we define the two actions that we shall 
consider in the manuscript. We study their finiteness and differentiability. In Sec.~\ref{sec:4} we employ 
the covariant Hamiltonian formalism to find the symplectic structure and the corresponding conserved 
quantities. In particular, we find an expression for the energy (up to a constant). In Sec.~\ref{sec:5} we 
perform the 2+1 decomposition of the action, following two different methods and obtain the energy as the 
contribution to the Hamiltonian coming from the boundary. We end with a discussion in Sec.~\ref{sec:6}. We 
have included two appendices.

Throughout the manuscript we set $c=1$, but leave the gravitational constant $G$ explicit. Note that we are {\emph{not}} setting $8\pi G=1$ as is normally done in the 2+1 literature.

\section{Preliminaries: Asymptotics in 3 dimensions}
\label{sec:2}

In this section we shall recall some subtleties that appear in the definition of asymptotically flat 3D 
spacetimes. We shall contrast the case at hand with that of ordinary 4D spacetimes.
Intuitively speaking, in $(3+1)$ dimensions we can think of an asymptotically flat spacetime as an 
spacetime with certain matter content in a bounded region outside of which the metric approaches the 
Minkowski metric. In the standard definition we say that a smooth space-time metric $g$ on 
$\mathcal{R}$ is \emph{weakly asymptotically flat  at spatial infinity} if there
exist a Minkowski metric $\eta$ such that, outside a spatially compact world tube, $(g - \eta)$ admits an 
asymptotic expansion to order 1 and $\lim_{r^{m} \rightarrow \infty } (g - \eta) = 0$.\footnote{The 
explicit form of the expansion depends on the coordinates. For instance, in 3-dimensions and cylindrical
coordinates, as we shall use through the present work, an asymptotic expansion to order $m$ of a function 
$f$ has the form,
    \begin{equation}
f(r, \theta) = \sum_{n=0} ^{m} \frac{\,^{n}f(\theta)}{r^{n}} + o(r^{-m}),
\end{equation}
where $r$ and $\theta$ are the coordinates on cylinders with $r = const$ and the remainder $o(r ^{-m})$ 
has the property that
\begin{equation}
\lim_{r \rightarrow \infty} r \,\,\, o(r ^{-m}) = 0.
\end{equation}}

In a (2+1) spacetime the situation is slightly different. For illustrative purposes, let us consider a mass distribution, say a point particle at the origin, $r=0$. Outside this region, $r>0$, the metric does not approach a flat metric, \emph{it is flat.} So, how can we define an asymptotically flat space-time?
In order to define an (2+1) asymptotically flat spacetime, we can first study this particular spacetime corresponding to a point particle of mass $M$ at the origin,
\begin{equation}
\d s^2=-\d t^2+r^{-8GM}(\d r^2+r^2\d\theta^2) \ \ \ {\mathrm{for}} \ r > 0 \label{pp}
\end{equation}
where $t,r,\theta$ are the cylindrical coordinates, $t\in(-\infty, +\infty)$, $r\in [0,\infty)$, and $\theta\in [0,2\pi)$. This metric is flat everywhere except at the origin. To see that, we can define $\rho:= \frac{r^\alpha}{\alpha}$, $\bar\theta:=\alpha\theta$ with $\alpha := 1-4GM$.
So the metric takes the form,
\begin{equation}
\d s^2=-\d t^2+\d\rho^2+\rho^2\d\bar\theta^2, \label{ppfl}
\end{equation}
from which the flatness of the metric is explicit. This is due to the fact that in a three-dimensional manifold satisfying Einstein's equations, whenever $T_{ab}=0$ the Riemann tensor is zero, i.e. the spacetime is flat on those points\footnote{We know that the Riemann tensor can be split into its trace and trace-free part, the Ricci tensor and scalar, and the Weyl tensor respectively. In 3-dimensions the Weyl tensor is identically zero, and by Einstein's equations if $T_{ab}=0$ implies that the Ricci tensor and scalar are also zero. Therefore the Riemann tensor is zero, so locally the space-time is flat. Note also that here we are dealing with asymptotically flat space-time, in contrast to the conformally flat picture where the vanishing of the Cotton tensor is equivalent to the metric being conformally flat.}.

In order to further understand the global structure of this spacetime, one can note that $\bar\theta \in 
[0, 2 \pi\alpha)$ with ($0 < \alpha \leq 1$). Therefore, there is a deficit angle which, despite 
the local flatness for $r>0$, makes this spacetime not globally equivalent to Minkowski space 
(due to the conic singularity). 

We are now in position of specifying the notion of asymptotic flatness for 3D gravity. Instead of 
requiring that all metrics approach a `single' Minkowski metric at infinity, one has now a one parameter 
family of possible, inequivalent, asymptotic configurations labelled, intuitively, by the ``mass $M$" of 
the asymptotic spacetime. That is,
we are looking for a metric that at spatial infinity approaches that of a point particle at 
the origin (\ref{pp}). 
Thus, we can define a 2+1 space-time to be asymptotically flat if, the line element admits an expansion of the form\footnote{A word on notation, ${\cal O} (r^{-m} )$ means that those terms \emph{include} a term proportional to $r^{-m}$ and terms that decay faster, in contrast with $o(r^{-m})$ that only includes terms that decay faster than $r^{-m}$, for instance, terms of the form $\frac{f}{r^{-m+\epsilon}}$.} \cite{Marolf-Patino2006},
\begin{eqnarray}\label{2mas1falloffmetric}
\nonumber \d s^2&=&-\left(1 +{\cal O}\left(\frac{1}{r}\right)\right)\d t^2  +r^{-\beta} \left[\left(1 +{\cal O}\left(\frac{1}{r}\right)\right)\d r^2+r^2\left(1 +{\cal O}\left(\frac{1}{r}\right)\right) \d\theta^2\right] \\
&&+{\cal O}(r^{-1-\beta/2})\d t\d\theta , 
\end{eqnarray}

Note that in the asymptotic region (when $r \rightarrow \infty$) the previous line element approaches to the background metric (in Cartesian coordinates),

\begin{equation}\label{eta}
\bar{\eta}_{ab} = \begin{pmatrix}
 -1 & 0 & 0  \\
  0 & r^{-\beta} & 0  \\
  0  & 0  & r^{-\beta}\\
 \end{pmatrix}. 
 \end{equation}

Note that we are approaching spatial infinity by some one-parameter family of boundaries of regions $\mathcal{M}_{\rho} \subset \mathcal{M}$ (cylinders throughout the present work, since they are more suited for Hamiltonian methods, as we plan to use in the following sections. Furthermore, the use of hyperboloids in the $3D$ context is less natural than in the $4D$ case \cite{aes, cwe, crv1,crv2}, due to the lack of asympototic Lorentz invariance, since, unless $M=0$, the asymptotically flat spacetime previously defined is not globally isometric to the three dimensional Minkowski space). $\{\mathcal{M}_{\rho} | \rho > 0\}$ are an increasing family, i.e. $\mathcal{M}_{\rho} \subset \mathcal{M}_{\rho '}$ whenever $\rho < \rho '$ and such that they cover $\mathcal{M}$ ($\bigcup_{\rho} \mathcal{M}_{\rho} = \mathcal{M}$). This procedure of taking a finite region $\mathcal{M}_{\rho}$ represents a cut-off for space-time and then we remove it by the limiting process $\rho \rightarrow \infty$. We take $\rho = r + {\cal O}(r^{0})$. This is called a `cylindrical cut-off' in \cite{mm}.

To summarize, an asymptotically flat spacetime  approaches that of a point particle (as opposed to a fixed Minkowski metric in the $4D$ case). In terms of the matter fields that might be present in the spacetime, the particular falloff conditions in the geometric degrees of freedom imply certain decay rates for matter. Since they do not have much of an impact in the quantities we are considering here (just as in the 3+1 case) we shall not consider any matter content in particular.  For a related treatment of asymptotically flatness from the conformal perspective (where the particular decay rates on matter are discussed), see \cite{symmetry}.

\section{The action and the boundary conditions of the first order variables}
\label{sec:3}

We can consider the Palatini action in three dimensions, whose equations of motion are equivalent to those given by the three dimensional Einstein-Hilbert action. Now the dynamical variables instead of the metric are a triad $e$ and a Lorentz connection $\omega$, both valued on the Lie algebra of $SO(2,1)$\footnote{The co-tetrad $e_a^I$ has an internal index $I$ `living' in an internal 3 dimensional vector space. Since the Lie algebra of $SO(2,1)$ is three dimensional, we can identify them.}.
Furthermore, we add to the Palatini action a boundary term in order to have a well posed action principle, that is, we want the action to be finite when evaluated on histories compatible with the boundary conditions, and also differentiable.\footnote{For further discussion on what it means for an action to be differentiable see \cite{crv1,crv2}.} 

As we have emphasized, we want to begin with a well posed action principle, so it is natural to start with the three dimensional analogue of the four dimensional well posed  Palatini action \cite{aes}. That is, let us define the  \emph{the Standard Palatini action with boundary term} (SPB) as,
\begin{equation}\label{SPB}
S_{\textrm{SPB}}[e,\omega] = - \frac{1}{ \kappa} \int_{\mathcal{M}} e^{I} \wedge F_{I} \; \; - \frac{1}{ \kappa} \int_{\partial \mathcal{M}} e^{I} \wedge \omega_{I}\, ,
\end{equation}
where $\kappa=8\pi G$.
Now, the natural question arises: is the boundary term gauge invariant? (under local Lorentz transformations). We can answer this in two ways. The first is by noting that we can perform a Lorentz transformation on the internal indices in (\ref{AFfalloff-cotriad}), (\ref{AFfalloff-triad}) and we still have an asymptotically flat configuration. So, in a sense, the internal directions are `arbitrary', therefore without loss of generality we can fix on the boundary one of the internal directions $\partial_{a} n^{I} = 0$ as in the 4-dimensional case \cite{aes,bn}, and the boundary term will be invariant under the residual gauge transformations. One should also expect that, just as in the 3+1 case  one has to fix the asymptotic tetrad in order to have a consistent formalism \cite{CR}, in our case this is also needed.

On the other hand we can add the following term to the action,
\begin{equation}\label{AdditionalHarmlessBoundaryTerm}
\frac{\alpha}{\kappa} \int_{\partial \mathcal{M} } \frac{1}{ n \cdot n} \varepsilon^{IKL} e_{I} \wedge n_{K} \mathrm{d} n_{L}
\end{equation}
with this addition, when $\alpha = 1$, the boundary term in (\ref{SPB}) becomes\footnote{
\begin{eqnarray}
\nonumber \int_{\partial \mathcal{M} } \frac{1}{n \cdot n} \varepsilon^{IKL} e_{I} \wedge n_{K} \mathcal{D} n_{L} &=& \int_{\partial \mathcal{M} } \frac{1}{n \cdot n} \varepsilon^{IK}\, _{L} e_{I} \wedge n_{K} \left(\mathrm{d}n^{K} + \varepsilon^{L}\, _{MN} \omega^{M} n^{N} \right)\\
&=& \int_{\partial \mathcal{M} } \frac{1}{n \cdot n} \varepsilon^{IKL} e_{I} \wedge n_{K} \mathrm{d}n_{L} + \int_{\partial \mathcal{M} } \frac{1}{n \cdot n} \varepsilon^{IKL} e_{I} \wedge n_{K} \varepsilon_{LMN} \omega^{M} n^{N}
\end{eqnarray}
},
\begin{equation}\label{AppLI-Result}
-\frac{1}{\kappa} \int_{\partial \mathcal{M}} e^{I} \wedge \omega_{I} -\frac{1}{\kappa} \int_{\partial \mathcal{M} } \frac{1}{ n \cdot n} \varepsilon^{IKL} e_{I} \wedge n_{K} \mathrm{d} n_{L} = -\frac{1}{\kappa} \int_{\partial \mathcal{M} } \frac{1}{n \cdot n} \varepsilon^{IKL} e_{I} \wedge n_{K} \mathcal{D} n_{L} .
\end{equation}
So instead of the action (\ref{SPB}) we can begin with the \emph{manifestly Lorentz invariant well posed action} (LIP)\footnote{Note the global minus sign, this is introduced since the Einstein Hilbert action with Gibbons Hawking term is equivalent to this action with minus sign (see appendix \ref{Appendix-Equiv-Einstein-vs-Palatini} for more details), so we can compare our results here with those obtained in the second order formulation \cite{Ashtekar-Varadarajan1994, Marolf-Patino2006}.
},
\begin{equation}\label{LIP}
S_{\textrm{LIP}}[e,\omega] = - \frac{1}{ \kappa} \int_{\mathcal{M}} e^{I} \wedge F_{I} \; \; - \frac{1}{ \kappa} \int_{\partial \mathcal{M} } \frac{1}{n \cdot n} \varepsilon^{IKL} e_{I} \wedge n_{K} \mathcal{D} n_{L}.
\end{equation}
Note that the \emph{general Palatini action} contains both the SPB and LIP cases, when $\alpha = 0$ and $\alpha = 1$ respectively, we shall use it to compare both actions,
\begin{equation}
S_{\textrm{GP}}[e,\omega]  = - \frac{1}{ \kappa} \int_{\mathcal{M}} e^{I} \wedge F_{I} \; \; - \frac{1}{ \kappa} \int_{\partial \mathcal{M}} e^{I} \wedge \omega_{I} - \frac{\alpha}{\kappa} \int_{\partial \mathcal{M} } \frac{1}{ n \cdot n} \varepsilon^{IKL} e_{I} \wedge n_{K} \mathrm{d} n_{L}
\end{equation}
Moreover, we can show that (\ref{AdditionalHarmlessBoundaryTerm}) is a constant when evaluated on 
asymptotically flat boundary conditions (see Appendix \ref{NewBoundaryTerm-constant} for the details on 
the derivation), so it does not spoil finiteness nor differentiability of the action. Therefore 
(\ref{LIP}) is still a well posed action. Further, the term (\ref{AppLI-Result}) is related the Gibbons-
Hawking term needed for the Einstein-Hilbert action to be well posed  and the action (\ref{LIP}) is 
\emph{the same} as the Einstein-Hilbert action with Gibbons-Hawking term \cite{Marolf-Patino2006}.

As in the four dimensional case this is a first order action, we only have first derivatives on our 
configuration variables, that is why we also refer to these variables as first order variables.

Some comments are in order. We are writing the action in a way that is independent of the Lie group $G$ on which is defined \cite{romano}, which does not need the existence of a metric to be defined. In the case of an arbitrary $G$, $e_{aI}$ can no longer be thought of as the cotriad. The action (\ref{SPB}) is then a functional of a $\pounds_{G}-$valued connection one-form $\omega_{a} ^{I}$ and a $\pounds_{G} ^{*}-$valued covector field $e_{aI}$. Where $\pounds_{G}-$ stands out for the Lie algebra of $G$ and $\pounds ^{\star} _{G}-$ its dual. When we chose $G=SO(2,1)$ we recover three-dimensional general relativity and we can think of $e_{aI}$ as a cotriad. This coincidence is exclusive of the three-dimensional case.\\

\subsection{Fall-off conditions}\label{SecFallOff}

To check that, in fact, the previous action is well posed we need to specify the boundary conditions on the first order variables $e$ and $\omega$, in this case asymptotically flat boundary conditions. 

From the line element (\ref{2mas1falloffmetric}),
\begin{eqnarray} 
\nonumber \d s^2&=&-\left(1 +{\cal O}\left(\frac{1}{r}\right)\right)\d t^2  +r^{-\beta} \left[\left(1 +{\cal O}\left(\frac{1}{r}\right)\right)\d r^2+r^2\left(1 +{\cal O}\left(\frac{1}{r}\right)\right) \d\theta^2\right] \\
&&+{\cal O}(r^{-1-\beta/2})\d t\d\theta , 
\end{eqnarray}
we can find the fall-off conditions of $g_{ab}$ as in \cite{Ashtekar-Varadarajan1994, Marolf-Patino2006}, with $a,b,c = 0,1,2$ spacetime indices, and therefore remembering that $g_{ab} = \eta_{IJ} e_{a} ^{I} e_{b} ^{J}$ where $\eta_{IJ} = \mathrm{diag}(-1,1,1)$ is the Minkowski metric, the fall-off conditions of the first order variables. 

We can assume that the co-triads and the triads admit an asymptotic expansion of the form\footnote{A tensor field $T^{a...b}\,_{c...d}$ will be said to admit an asymptotic expansion to order $m$ if all its component in the \emph{Cartesian} chart $x^{a}$ do so. Note that apart from the $r^{-\beta}$ factor in the spatial part of (\ref{2mas1falloffmetric}) the components in cartesian coordinates admit an expansion of order 1 in analogy with the standard definition of an asymptotically flat spacetime for 4 dimensional spacetimes \cite{aes, cwe, crv1,crv2}, and also we assume that the first order variables, apart from a factor of $r^{-\beta /2}$, do so.}
\begin{equation}\label{AFfalloff-cotriad}
 e^{I} _{a} = \delta^{0} _{a} \left(\,^{o}\bar{e}^{I} _{0} + \frac{\,^{1}\bar{e}^{I} _{0}(\theta )}{r} + o(r^{-1})\right) + r^{-\beta / 2} \left(\,^{o}\bar{e}^{I} _{\bar{a}} + \frac{\,^{1}\bar{e}^{I} _{\bar{a}}(\theta )}{r} +  o(r^{-1})\right) \delta^{\bar{a}} _{a},
\end{equation}
and
\begin{equation}\label{AFfalloff-triad}
e_{I} ^{a} = \delta_{0} ^{a} \left(\,^{o}\bar{e}_{I} ^{0} + \frac{\,^{1}\bar{e}_{I} ^{0}(\theta )}{r} +  o(r^{-1})\right) + r^{\beta / 2} \left(\,^{o}\bar{e}_{I} ^{\bar{a}} + \frac{\,^{1}\bar{e}_{I} ^{\bar{a}}(\theta )}{r}  + o(r^{-1})\right) \delta_{\bar{a}} ^{a}.
\end{equation}
We define,
\begin{equation}\label{Def-0e}
\, ^{0} e_{a} ^{I} := \, ^{0} \bar{e}_{0} ^{I} \delta^{0} _{a} + r ^{- \beta / 2} \,^{0} \bar{e}_{\bar{a}} ^{I} \delta^{\bar{a}} _{a} \,\,\,\,\, \mathrm{and} \,\,\,\,\,  \, ^{1} e_{a} ^{I} := \frac{ \, ^{1} \bar{e}_{0} ^{I} }{r} \delta^{0} _{a} + r ^{- \beta / 2} \frac{\,^{1} \bar{e}_{\bar{a}} ^{I}}{r} \delta^{\bar{a}} _{a}
\end{equation}
such that $\bar{\eta} _{ab} = \eta_{IJ} \,^{0} e_{a} ^{I} \,^{0} e_{b} ^{J}$ given by (\ref{eta}), where $\eta_{IJ} = \mathrm{diag}(-1,1,1)$ is the Minkowski metric.

As for the triads, we assume that the connection $\omega_{a} ^{I}$ admits an expansion of the form,
\begin{equation}\label{AFfalloff-connection}
\omega^{I} _{a} =  \,^{o}\bar{\omega}^{I} _{a} + \frac{\,^{1}\bar{\omega}^{I} _{a}(\theta )}{r} +  \frac{\,^{2}\bar{\omega}^{I} _{a}(\theta )}{r^{2}} + o(r^{-2}),
\end{equation}
Even though this expansion seems different from that of the triad, we can check that this expansion is derived from that of the triad and co-triad by means of the condition, De = 0, to first order.

Now we have to recall that any connection $D$ can be written as $D = \mathring{\bar{\mathcal{D}}} + \omega$ , where $\mathring{\bar{\mathcal{D}}}$ is any other connection. When there is a `preferred' connection available, we can write all the other connections as that one plus a vector potential $\omega$. Since there is no canonical choice of this \emph{standard flat connection}, $\mathring{\bar{\mathcal{D}}}$, within this particular problem it will be convenient to choose that $\mathring{\bar{\mathcal{D}}}_{[a} \, ^{0} \bar{e}_{b]} ^{I} = 0$. Using local coordinates and a local trivialization of $E = U_\mathcal{M} \times SO(2,1)$, where $U_\mathcal{M}$ is an open set on $\mathcal{M}$, the components of the connection for the condition of the compatibility of the triad with the connection, $De = 0$ will look like,
\begin{equation}\label{Deigual0}
D_{[a} e_{b]} ^{I} = \mathring{\bar{\mathcal{D}}}_{[a} e_{b]} ^{I} + \varepsilon^{IJK} \omega_{[a|J} e_{b]K} = 0.
\end{equation} 

From (\ref{Deigual0}) it is a straightforward calculation to see that the spin connection can be written in terms of the triad as,
\begin{equation}
\omega_{c} ^{M} = - \frac{1}{2} \left( \varepsilon_{L} \, ^{KM} e^{a} _{K} e^{bL} e_{cI} \mathring{\bar{\mathcal{D}}}_{[a} e_{b]} ^{I} - \varepsilon_{L} \, ^{KM} e^{a} _{K} \mathring{\bar{\mathcal{D}}}_{[c} e_{a]} ^{L} - \varepsilon_{L} \, ^{KM} e^{bL} \mathring{\bar{\mathcal{D}}}_{[b} e_{c] K}  \right).
\end{equation}

The leading term of the spin connection can be found from the previous equation considering the leading terms of the triad and cotriad,
\begin{equation}\label{zeroOmegaExpressionLeadingterm}
 \, ^{Leading} \omega_{c} ^{M} = - \frac{1}{2} \left( \varepsilon_{L} \, ^{KM} \,^{0} e^{a} _{K} \,^{0} e^{bL}\,^{0} e_{cI} \mathring{\bar{\mathcal{D}}}_{[a} \,^{0} e_{b]} ^{I} - \varepsilon_{L} \, ^{KM} \,^{0} e^{a} _{K} \mathring{\bar{\mathcal{D}}}_{[c} \,^{0} e_{a]} ^{L} - \varepsilon_{L} \, ^{KM} \,^{0} e^{bL} \mathring{\bar{\mathcal{D}}}_{[b} \,^{0} e_{c] K}  \right). 
\end{equation}
where $\mathring{\bar{\mathcal{D}}}_{b} \, ^{0} \bar{e}_{a} ^{I} = 0$. Note that from (\ref{AFfalloff-cotriad}), 
\begin{equation}
\mathring{\bar{\mathcal{D}}}_{b} \, ^{0} e_{a} ^{I} = \mathring{\bar{\mathcal{D}}}_{b} ( \, ^{0}  \bar{e}_{0} ^{I} \delta_{a} ^{0} ) + \mathring{\bar{\mathcal{D}}}_{b} (r^{-\beta /2} \,^{0} \bar{e}_{\bar{a}} ^{I} \delta_{a} ^{\bar{a}}  ) = \mathring{\bar{\mathcal{D}}}_{b} (r^{-\beta /2}) \,^{0} \bar{e}_{\bar{a}} ^{I} \delta_{a} ^{\bar{a}}  = (\partial_{b} r^{- \beta / 2}) \,^{0} \bar{e}_{\bar{a}} ^{I} \delta_{a} ^{\bar{a}}
\end{equation}
but $\partial_{b} r^{- \beta / 2} = - \frac{1}{2} \beta r^{-1 - \beta / 2} \partial_{b} r$. Therefore,
\begin{equation}\label{0barDe}
\mathring{\bar{\mathcal{D}}}_{b} \, ^{0} e_{a} ^{I} = (- \frac{1}{2} \beta r^{-1 - \beta / 2} \partial_{b} r) \,^{0} \bar{e}_{\bar{a}} ^{I} \delta_{a} ^{\bar{a}}= (- \frac{1}{2} \beta r^{-1 } \partial_{b} r) \,^{0} e_{\bar{a}} ^{I} \delta_{a} ^{\bar{a}}
\end{equation}
Taking into account the previous equation and the fall-off conditions (\ref{AFfalloff-cotriad}) and (\ref{AFfalloff-triad}), equation (\ref{zeroOmegaExpressionLeadingterm}) becomes (using that $\partial_{0} r = 0$),
\begin{equation}
 \, ^{Leading} \omega_{c} ^{M} = \frac{\beta}{2r} \varepsilon_{L} \, ^{KM} \,^{0} \bar{e}_{K} ^{\bar{a}} \,^{0} \bar{e}_{\bar{c}} ^{L} \delta_{c} ^{\bar{c}},
\end{equation}
then considering the expansion (\ref{AFfalloff-connection}) we can see that,
\begin{equation}\label{FalloffOmegaZero}
\,^{1} \bar{\omega}_{c} ^{M} =  \frac{\beta}{2}  \partial_{\bar{a}} r \varepsilon_{L}\, ^{KM} \,^{0} \bar{e}_{K} ^{\bar{a}} \,^{0} \bar{e}_{\bar{c}} ^{L} \delta_{c} ^{\bar{c}}.
\end{equation}
Which implies that $\frac{\,^{1} \bar{\omega}_{c} ^{M}}{r}$ is the leading term of $\omega_{c} ^{M}$ and that $\,^{0} \omega_{c} ^{M} = 0$ as well as $\,^{1} \omega_{0} ^{M} = 0$.

\subsection{Well posedness of the action}

As we already mentioned, beginning with a well posed action principle under asymptotically flat boundary conditions, we want to find an expression for the energy under various approaches. We want to analyse whether this results coincide with those in the second order formalism \cite{Ashtekar-Varadarajan1994, Marolf-Patino2006} and also the relation and differences among the different paths we take: the covariant Hamiltonian formalism (CHF), and the canonical one, where we take two different $2+1-$decompositions.

But first we have to check that the action principle we are working with is well posed, i.e. finite and differentiable under asymptotically flat boundary conditions and variations. With the fall-off conditions of the first order variables found in section \ref{SecFallOff} we are ready to undertake this task.

\subsubsection{Finiteness}\label{Finiteness}

Since the term (\ref{AdditionalHarmlessBoundaryTerm}) is a finite constant when evaluated on the boundary\footnote{See appendix \ref{NewBoundaryTerm-constant} for details.}, it does not spoil finiteness. Then, it is only necessary to cheek that the action (\ref{SPB}) is finite, so the manifestly gauge invariant action (\ref{LIP}) is also finite. The action  (\ref{SPB}) can be rewritten as,

\begin{eqnarray}
\nonumber S[e, \omega]  &=& - \frac{1}{ \kappa} \int_{\mathcal{M}} e^{I} \wedge F_{I} \; \; - \frac{1}{ \kappa} \int_{\partial \mathcal{M}} e^{I} \wedge \omega_{I}\\
&=& - \frac{1}{ \kappa} \int_{\mathcal{M}} \left(e^{I} \wedge \mathrm{d} \omega_{I} + \frac{1}{2} \varepsilon_{I} \, ^{JK} e^{I} \wedge \omega_{J} \wedge \omega_{K}  \right) \; \; - \frac{1}{ \kappa} \int_{\partial \mathcal{M}} e^{I} \wedge \omega_{I}
\end{eqnarray}
since $F_{I} = \mathrm{d} \omega_{I} + \frac{1}{2} \varepsilon_{I} \, ^{JK} \omega_{J} \wedge \omega_{K}  $ and,
\begin{equation}
\mathrm{d} (e^{I} \wedge \omega_{I} ) = \mathrm{d} e^{I} \wedge \omega_{I} - e^{I} \wedge \mathrm{d} \omega_{I} \,\, \Rightarrow \,\, e^{I} \wedge \mathrm{d} \omega_{I} = \mathrm{d} e^{I} \wedge \omega_{I} - \mathrm{d} (e^{I} \wedge \omega_{I} ).
\end{equation}
Then,
\begin{eqnarray}\label{ExpressionFiniteness}
\nonumber S[e, \omega]  &=& -\frac{1}{ \kappa}  \int_{\mathcal{M}} \left(\mathrm{d} e^{I} \wedge \omega_{I} + \frac{1}{2} \varepsilon_{I} \, ^{JK} e^{I} \wedge \omega_{J} \wedge \omega_{K} - \mathrm{d} (e^{I} \wedge \omega_{I} )\right) \; \; -  \frac{1}{ \kappa}\int_{\partial \mathcal{M}} e^{I} \wedge \omega_{I}\\
&=& - \frac{1}{ \kappa} \int_{\mathcal{M}} \left(\mathrm{d} e^{I} \wedge \omega_{I} + \frac{1}{2} \varepsilon_{I} \, ^{JK} e^{I} \wedge \omega_{J} \wedge \omega_{K} \right).
\end{eqnarray}

The leading term of the previous equation is,
\begin{equation}\label{ActionFinitenessFirstorder}
\,^{0} S[e, \omega] = - \frac{1}{ \kappa} \int_{\mathcal{M}} \left(\mathrm{d} \,^{0} e^{I} \wedge \,^{1} \omega_{I} + \frac{1}{2} \varepsilon_{I} \, ^{JK} \,^{0} e^{I} \wedge \,^{1} \omega_{J} \wedge \,^{1} \omega_{K} \right),
\end{equation}
but we already used the compatibility condition with the triad to first order to obtain the fall-off conditions on $\omega$, (\ref{Deigual0}), which can also be written as,
\begin{equation}
\mathrm{d} \,^{0} e^{I} - \varepsilon^{I}\, _{JK} \,^{1} \omega^{K} \wedge \,^{0} e^{J} = 0
\end{equation}
therefore, we can rewrite  (\ref{ActionFinitenessFirstorder}) as,
\begin{eqnarray}
\nonumber \,^{0} S[e, \omega]  &=& - \frac{1}{ \kappa} \int_{\mathcal{M}} \left(\mathrm{d} \,^{0} e^{I} \wedge \,^{1} \omega_{I} - \frac{1}{2} \varepsilon_{I} \, ^{JK} \,^{0} e^{I} \wedge  \,^{1} \omega_{J} \wedge \,^{1} \omega_{K} \right) \\
&=& -  \frac{1}{ \kappa} \int_{\mathcal{M}} \left( \mathrm{d} \,^{0} e^{I} \wedge \,^{1} \omega_{I} - \frac{1}{2} \mathrm{d} \,^{0} e^{I} \wedge \,^{1} \omega_{I} \right) = - \frac{1}{ \kappa} \int_{\mathcal{M}} \frac{1}{2} \mathrm{d} \,^{0} e^{I} \wedge \,^{1} \omega_{I}.
\end{eqnarray}
Now, using (\ref{0barDe}) and (\ref{FalloffOmegaZero}) the leading term is\footnote{Where $\d x^{a} \wedge \d x^{b} \wedge \d x^{c} = \tilde{\varepsilon}^{abc}\d^{3}x$, with $\tilde{\varepsilon}^{abc}$ the Levi-Civita tensor density of weight +1, that is related with the Levi-Civita tensor, $\varepsilon^{abc}$, by $\tilde{\varepsilon}^{abc} = (s) \sqrt{|g|} \varepsilon^{abc}$ with $g$ the determinant of the spacetime metric and $s$ the signature of the metric. },
\begin{equation}\label{Finiteness-LeadingZero}
\frac{1}{4\kappa} \int_{\mathcal{M}}  \mathring{\bar{\mathcal{D}}}_{a} \, ^{0} e_{b} ^{I} \,^{1} \omega_{c} ^{K} \tilde{\varepsilon}^{abc} \d^{3} x = 0,
\end{equation}
since\footnote{$\,^{1} \omega_{0} ^{K} = 0$ is zero from the fall off conditions on $\omega$, $\mathring{\bar{\mathcal{D}}}_{a} \, ^{0} e_{0} ^{I} = 0$ because $\, ^{0} e_{0} ^{I} = \, ^{0} \bar{e}_{0} ^{I}$ and $D_{0} \, ^{0} e_{\bar{a}} ^{I} = 0$ because we ask the condition of the compatibility of the triad with the connection to be satisfied to first order to find the fall-off conditions on $\omega$,
\[ D_{0} \,^{0} e_{b} ^{I} = \mathring{\bar{\mathcal{D}}}_{0} \,^{0} e_{b} ^{I} + \varepsilon^{IJK} \,^{1} \omega_{0J} \,^{0} e_{bK} = 0.
\]
since $\,^{1} \omega_{0} ^{K} = 0$ then $\mathring{\bar{\mathcal{D}}}_{0} \,^{0} e_{b} ^{I} =0$. } $\,^{1} \omega_{0} ^{K} = 0$, $\mathring{\bar{\mathcal{D}}}_{0} \, ^{0} e_{\bar{a}} ^{I} = 0$ and $\mathring{\bar{\mathcal{D}}}_{a} \, ^{0} e_{0} ^{I} = 0$. On the other hand note that we could have chosen to write (\ref{ExpressionFiniteness}), using $De =0$ to first order as well, as,
\begin{equation}
\,^{0} S[e, \omega]   = - \frac{1}{4 \kappa}   \int_{\mathcal{M}} \varepsilon^{I}\, _{JK} \,^{0} e^{I} \wedge \,^{1} \omega^{J} \wedge \,^{1} \omega^{K} = - \frac{1}{4 \kappa} \int_{\mathcal{M}} \varepsilon^{I}\, _{JK} \,^{0} e_{a} ^{I} \,^{1} \omega_{b} ^{J} \,^{1} \omega_{c} ^{K} \tilde{\varepsilon}^{abc} \d^{3} x.
\end{equation}
In the previous equation, using (\ref{Def-0e}) and (\ref{FalloffOmegaZero}), the only nonvanishing term is
\begin{equation}
\,^{0} S[e, \omega]  = - \frac{1}{4 \kappa} \int_{\mathcal{M}} \varepsilon^{I}\, _{JK} \,^{0} \bar{e}_{0} ^{I} \frac{ \,^{1} \bar{\omega}_{\bar{b}} ^{J}}{r} \frac{\,^{1} \bar{\omega}_{\bar{c}} ^{K}}{r}  \varepsilon^{0\bar{b}\bar{c}}\, r \d r \d \theta \d t =  \int_{\mathcal{M}}{ \cal{O}}( r^{-1}) \d r\, .
\end{equation}

Our region of integration $\mathcal{M}$ is bounded by $\partial \mathcal{M} = M_{1} \cup M_{2} \cup \mathcal{I}$ with its corresponding orientation. In order to check finiteness it is enough to check that the integral over a spatial hypersurface is finite. This is true since we are integrating over a finite time interval where the Cauchy surfaces $M_{1}$ and $M_{2}$ are asymptotically time-translated with respect to each other. Such spacetimes $\mathcal{M}$ are referred to as \emph{cylindrical slabs} \cite{aes} or as \emph{cylindrical temporal cut-off} \cite{mm}.

Note that on a Cauchy slice the only dependency on $r$ of the previous equation  is due to $\,^{1} 
\omega_{c}^{K} ={ \cal{O}}( r^{-1})$, so the integral over $r$ goes as $\int { \cal{O}}( r^{-1}) \d r$ that 
\emph{may} logarithmically diverge in the limit $r \rightarrow \infty$, but we already proved in 
(\ref{Finiteness-LeadingZero}) that this term is zero. Then, the next to leading terms decay faster in $r$ 
so, in the limit $r \rightarrow \infty$, they go to zero. 
Therefore, the integral is finite \emph{even off shell}.

\subsubsection{Differentiability}

In order for an action to be differentiable the variation of the action needs to take the form,
\begin{equation}
\delta S [e,\omega] =  \int_{\mathcal{M}} \left[\mathbf{E}_{e} \wedge \delta e   +  \mathbf{E}_{\omega} \wedge \delta \omega \right] +  \int_{\partial M} \tilde{\theta} (e^{I}, \omega^{I}, \delta e^{I}, \delta \omega^{I}),
\end{equation}
and in order for $\mathbf{E}_{e}$ and $\mathbf{E}_{\omega}$ to be the Euler-Lagrange equations of motion, the boundary term needs to be zero when evaluated on histories compatible with the boundary conditions. Since the term (\ref{AdditionalHarmlessBoundaryTerm}) is constant when evaluated on those histories, its variation is zero so it does not spoil differentiability. Therefore we only need to check whether the action (\ref{SPB}) is differentiable.

The variation of the 3-dimensional  Palatini action with boundary term (\ref{SPB}) is, 
\begin{equation}
\delta S [e,\omega] = - \frac{1}{ \kappa} \int_{\mathcal{M}} \left[\delta e^{I} \wedge F_{I}  + e^{I} \wedge \delta F_{I}\right] - \frac{1}{ \kappa} \int_{\partial \mathcal{M}} \left[\delta e^{I} \wedge \omega_{I}  + e^{I} \wedge \delta \omega_{I}\right],
\end{equation}
but
\begin{equation}
\delta F_{I} = \mathrm{d} \delta \omega_{I} + \frac{1}{2} \varepsilon_{I}\, ^{JK} \delta \omega_{J} \wedge \omega_{K} + \frac{1}{2} \varepsilon_{I} \, ^{JK} \omega_{J} \wedge \delta \omega_{K} = \mathrm{d} \delta \omega_{I} + \varepsilon_{I}\, ^{JK} \delta \omega_{J} \wedge \omega_{K}
\end{equation}
then, the variation becomes,
\begin{eqnarray}\label{variation}
\delta S [e,\omega] = - \frac{1}{ \kappa} \int_{\mathcal{M}} \delta e^{I} \wedge F_{I}  - \frac{1}{ \kappa} \int_{\mathcal{M}}  \left( \mathrm{d} e^{J} + \varepsilon^{JIK} e_{I} \wedge \omega_{K} \right) \wedge \delta \omega_{J} - \frac{1}{ \kappa} \int_{\partial \mathcal{M}} \delta e^{I} \wedge \omega_{I}  .
\end{eqnarray}
If the boundary term is zero under the boundary conditions, the action is said to be differentiable and the equations of motion are,
\begin{equation}\label{PalatiniEoM}
F_{I} = 0 \,\,\,\,\, \mathrm{and} \,\,\,\,\, De^{J} = \mathrm{d} e^{J} + \varepsilon^{JIK} e_{I} \wedge \omega_{K} = 0.
\end{equation}

That are equivalent to those given by the three-dimensional Einstein-Hilbert action. The boundary term is,
\begin{equation}
- \frac{1}{ \kappa}\int_{\partial \mathcal{M}} \delta e^{I} \wedge \omega_{I}  = - \frac{1}{ \kappa} \left( -\int_{M_{1}} + \int_{M_{2}} + \int_{\mathcal{I}} \right) \delta e^{I} \wedge \omega_{I} 
\end{equation}
where we are considering that our integration region $\mathcal{M}$ is bounded by $\partial \mathcal{M} = M_{1} \cup M_{2} \cup \mathcal{I}$ with its corresponding orientation. We are taking, as usual, $\delta e^{I} = \delta \omega_{I} = 0$ on the space-like surfaces $M_{1}$ and $M_{2}$. We are left only with the integral on the time-like boundary $\mathcal{I}$. Recall that we are approaching spatial infinity by a family of cylinders, $C_{r}$ with $r=const$, in the limit when $r \rightarrow \infty$.
To check differentiability  we have to prove that
\begin{equation}\label{BoundaryTerm-differentiability}
\lim_{r \rightarrow \infty} \int_{C_{r}} \delta e^{I} \wedge \omega_{I} = 0,
\end{equation}
when evaluated on histories compatible with the asymptotically flat boundary conditions. Note that we are allowing all the possible variations compatible with the boundary conditions and not only those of compact support. It is enough to check the behaviour of the leading term (the next to leading terms decay `faster' as $r$ goes to infinity). 
Considering the asymptotic conditions on $e_{a} ^{I}$ and $\omega_{a} ^{I}$, (\ref{AFfalloff-cotriad}) and (\ref{AFfalloff-connection}), and the fact that $\frac{ \,^{1} \bar{\omega}_{a} ^{I}}{r}$ is the leading term of $\omega_{a} ^{I}$ (thus $\,^{0}\omega_{a} ^{I} = 0$) with $\,^{1}\omega_{0} ^{I} = 0$; and using (\ref{Def-0e}), equation (\ref{BoundaryTerm-differentiability}) can be written as\footnote{Where $\varepsilon^{ab}$ is the two-dimensional Levi-Civita tensor related to the tensor density of weight +1 by $\varepsilon^{ab} = \frac{(s)}{\sqrt{|\gamma|}} \tilde{\varepsilon}^{ab}$, where $\gamma_{ab}$ is the induced metric on the timelike boundary, $\gamma$ its determinant and $s$ the signature of $\gamma_{ab}$.},
\begin{eqnarray}\label{BoundaryTerm-differentiability-Full}
\nonumber \lim_{r \rightarrow \infty} \int_{C_{r}} \delta e^{I} \wedge \omega_{I}  &=& \lim_{r \rightarrow \infty} \int_{C_{r}} \delta \left( \,^{0} e_{a} ^{I} + \,^{1} e_{a} ^{I} + o(r^{-2})  \right) \left( \frac{ \,^{1} \bar{\omega}_{\bar{b}} ^{J}}{r} \delta_{b} ^{\bar{b}}  + \frac{ \,^{2} \bar{\omega}_{b} ^{J}}{r^{2}} + o(r^{-2})  \right) \eta_{JI} \varepsilon^{ab} r \d \theta \d t\\
\nonumber &=& \lim_{r \rightarrow \infty} \int_{C_{r}} \left[ \delta ^{0} e_{a} ^{I} \,\, \frac{ \,^{1} \bar{\omega}_{\bar{b}} ^{J}}{r} \delta_{b} ^{\bar{b}}   +  \delta ^{0} e_{a} ^{I} \,\, \frac{ \,^{2} \bar{\omega}_{b} ^{J}}{r^{2}} + \delta ^{1} e_{a} ^{I} \,\, \frac{ \,^{1} \bar{\omega}_{\bar{b}} ^{J}}{r} \delta_{b} ^{\bar{b}} +  o(r^{-2})   \right] \eta_{JI} \varepsilon^{ab} r \d \theta \d t\\
&=& \lim_{r \rightarrow \infty} \int_{C_{r}} \left[ \delta ^{0} e_{a} ^{I} \,\, \frac{ \,^{1} \bar{\omega}_{\bar{b}} ^{J}}{r} \delta_{b} ^{\bar{b}}  + { \cal{O}}( r^{-2})\right] \eta_{JI} \varepsilon^{ab} r \d \theta \d t.
\end{eqnarray}
but\footnote{
From (\ref{Def-0e}) and since $\,^{0}\bar{e} _{a} ^{I}$ is a fixed flat frame at the asymptotic region, $\delta \,^{0}\bar{e} _{a} ^{I} = 0$, then,
\begin{equation}
\delta \,^{0} e_{a} ^{I} = \delta \left( \,^{0} \bar{e}_{0} ^{I} \delta_{a} ^{0} + r^{- \beta /2} \,^{0} \bar{e}_{\bar{a}} ^{I} \delta_{a} ^{\bar{a}}    \right) = \delta (r^{-\beta / 2}) \,^{0} \bar{e}_{\bar{a}} ^{I} \delta_{a} ^{\bar{a}}.
\end{equation}
In the timelike boundary $\delta r = 0$ so
\begin{eqnarray}
\nonumber \delta \left( r^{-\beta/2} \right) &=& - \frac{\beta }{2} r^{-\beta / 2 - 1} \delta r  - \frac{r^{- \beta / 2}}{2} \log (r) \delta \beta\\
&=& - \frac{r^{- \beta / 2}}{2} \log (r) \delta \beta
\end{eqnarray}}
\begin{equation}\label{delta-Ceroe}
\delta \,^{0} e_{a} ^{I} = \left( - \frac{r^{- \beta / 2}}{2} \log (r) \delta \beta \right) \,^{0} \bar{e}_{\bar{a}} ^{I} \delta_{a} ^{\bar{a}}.
\end{equation}
Using (\ref{delta-Ceroe}), equation (\ref{BoundaryTerm-differentiability-Full}) becomes,
\begin{eqnarray}\label{BoundaryTerm-differentiability-FullFull}
\nonumber \lim_{r \rightarrow \infty} \int_{C_{r}} \delta e^{I} \wedge \omega_{I} &=&  \lim_{r \rightarrow \infty} \int_{C_{r}} \left[ \left( - \frac{r^{- \beta / 2}}{2} \log (r) \delta \beta \right) \,^{0} \bar{e}_{\bar{a}} ^{I} \delta_{a} ^{\bar{a}} \,\, \frac{ \,^{1} \bar{\omega}_{\bar{b}} ^{J}}{r} \delta_{b} ^{\bar{b}}  + { \cal{O}}( r^{-2})\right] \eta_{JI} \varepsilon^{ab} r \d \theta \d t\\
&=&  \lim_{r \rightarrow \infty} \int_{C_{r}} \left[  - \frac{r^{- \beta / 2}}{2} \log (r) \delta \beta  \,\,^{0} \bar{e}_{\bar{a}} ^{I}   \,^{1} \bar{\omega}_{\bar{b}} ^{J} \delta_{a} ^{\bar{a}} \delta_{b} ^{\bar{b}}  + { \cal{O}}( r^{-1})\right] \eta_{JI} \varepsilon^{ab}  \d \theta \d t.
\end{eqnarray}
We can see in two ways that this term vanish. The first is to note that $\varepsilon^{ab}$ is the induced Levi-Civita tensor on the timelike boundary (hypercylinders) so the indices, $a,b=0,1$, have one temporal and one spatial component, but in the previous equation due to $\delta_{a} ^{\bar{a}} \delta_{b} ^{\bar{b}}\varepsilon^{ab} = 0$, the leading term vanishes identically. Also in the previous equation, we can note that the only dependence on $r$ is through $ r^{- \beta / 2} \log (r)$, and since we are not integrating over $r$ and demanding that $\beta >0$, 
\begin{equation}
 \lim_{r \rightarrow \infty} r^{- \beta / 2} \log (r) = 0.
\end{equation}
So in the limit equation (\ref{BoundaryTerm-differentiability-FullFull}) vanishes,
\begin{eqnarray}\label{DifferentiabilityProff}
\!\!\!\!\!\!\!\! \lim_{r \rightarrow \infty} \int_{C_{r}} \delta e^{I} \wedge \omega_{I} &=& \lim_{r \rightarrow \infty} \int_{C_{r}} \left[  - \frac{r^{- \beta / 2}}{2} \log (r) \delta \beta  \,\,^{0} \bar{e}_{\bar{a}} ^{I}   \,^{1} \bar{\omega}_{\bar{b}J} \delta_{a} ^{\bar{a}} \delta_{b} ^{\bar{b}}  + { \cal{O}}( r^{-1})\right] \varepsilon^{ab}  \d \theta \d t = 0.
\end{eqnarray}
Therefore the action is also differentiable under asymptotically flat boundary conditions, for arbitrary compatible variations.

\section{Covariant analysis}
\label{sec:4}

In this section we shall follow the approach of the covariant Hamiltonian formalism (CHF), as summarized in\cite{crv1,crv2}. In particular, we shall identify several components of the CHF, such as the symplectic potential, (pre-)symplectic structure and Hamiltonian generators, starting from the actions defined in Sec.~\ref{sec:3}. This section has two parts. In the first one we identify these quantities and prove their finiteness. In the second one we focus our attention on Hamiltonian flows and their generators.

\subsection{Symplectic geometry}

From the variation of the action (\ref{variation}), we can identify the symplectic potential,
\begin{equation}
\tilde{\Theta} (e^{I}, \omega^{I}, \delta e^{I}, \delta \omega^{I}) :=  \int_{\partial M} \tilde{\theta} (e^{I}, \omega^{I}, \delta e^{I}, \delta \omega^{I}) =  - \frac{1}{ \kappa} \int_{\partial M} \delta e^{I} \wedge \omega_{I},
\end{equation}
and its associated symplectic current,
\begin{equation}
J(\delta_{1}, \delta_{2}) :=  2 \delta_{[1} \tilde{\theta} ( \delta_{2]}) = -\frac{1}{ \kappa} \left( \delta_{2} e^{I} \wedge \delta_{1} \omega_{I} - \delta_{1} e^{I} \wedge \delta_{2} \omega_{I}\right).
\end{equation}
Since $J$ is closed over any region $\mathcal{M}$,
\begin{equation}
0 = \int_{\mathcal{M}} \mathrm{d} J (\delta_{1}, \delta_{2}) = \oint_{\partial \mathcal{M}} J ( \delta_{1}, \delta_{2}) = \left[ - \int_{M_{1}} + \int_{M_{2}} + \int_{\mathcal{I}} \right] J (\delta_{1}, \delta_{2}) 
\end{equation}
here we are  considering  the region $\mathcal{M}$ is bounded by $\partial_{\mathcal{M}} = M_{1} \cup M_{2} \cup \mathcal{I}$, $M_{1}$ and $M_{2}$ are space-like slices and $\mathcal{I}$ an outer boundary, in particular we shall consider configurations that are asymptotically flat. We are assuming no internal boundary.

In order to have a \emph{conserved symplectic current} and therefore a \emph{conserved pre-symplectic form}, independent of the Cauchy surface, we have to check that $\int_{\mathcal{I}} J = 0 $, that is, that there is no current `leakage' at infinity.

Taking into account the asymptotically flat boundary conditions previously derived, we can see that the leading terms of $\int_{\mathcal{I}} J$ are,
\begin{equation}\label{JatInfinity}
\int_{\mathcal{I}} \,^{0}  J (\delta_{1}, \delta_{2}) = - \frac{1}{ \kappa} \lim_{r \rightarrow \infty} \int_{C_{r}} \left( \delta_{2} \,^{0} e^{I} \wedge \delta_{1}  \,^{1} \omega_{I} - \delta_{1} \,^{0} e^{I} \wedge \delta_{2}  \,^{1} \omega_{I}  \right).
\end{equation}

Following the same arguments as in (\ref{DifferentiabilityProff}), that is using $\,^{1} \bar{\omega}_{0} ^{I} =0$ and $\delta \,^{0}e^{I} _{0} = 0$, and noticing that the previous equation becomes,
\begin{equation}
\int_{\mathcal{I}} \,^{0}  J (\delta_{1}, \delta_{2}) = - \frac{1}{ \kappa} \lim_{r \rightarrow \infty} \int_{C_{r}} \left( \delta_{2} \,^{0} e_{0} ^{I} \delta_{1}  \,^{1} \omega_{\bar{a}I} - \delta_{2} \,^{0} e_{\bar{a}} ^{I} \delta_{1}  \,^{1} \omega_{0I} - \delta_{1} \,^{0} e_{0}^{I}  \delta_{2}  \,^{1}   \omega_{\bar{a}I} +  \delta_{1} \,^{0} e_{\bar{a}} ^{I} \delta_{2}  \,^{1}   \omega_{0I}  \right) \tilde{\varepsilon}^{0 \bar{a}} \d^{2}x,
\end{equation}
we can see that 
\begin{equation}
\int_{\mathcal{I}} \,^{0}  J (\delta_{1}, \delta_{2}) = 0.
\end{equation}
But, on the other hand note that 
\begin{equation}
\int_{\mathcal{I}} \,^{0} J (\delta_{1}, \delta_{2}) = - \frac{1}{ \kappa} \lim_{r \rightarrow \infty} \int_{C_{r}} \left(  \delta_{2} \,^{0} e_{a} ^{I}  \delta_{1}  \frac{\,^{1} \bar{\omega}_{bI}}{r} - \delta_{1} \,^{0} e_{a} ^{I} \wedge \delta_{2}  \frac{\,^{1} \bar{\omega}_{bI}}{r}  \right) \varepsilon^{ab} r \d \theta \d t
\end{equation}
is independent of $r$. Therefore the next to leading terms goes as,
\begin{equation}
\int_{\mathcal{I}}  J (\delta_{1}, \delta_{2}) = - \frac{1}{ \kappa} \lim_{r \rightarrow \infty} \int_{C_{r}} {\cal O} (r^{-1} ) \varepsilon^{ab}  \d \theta \d t = 0.
\end{equation}
Therefore, the symplectic current is conserved.

Now we can define a \emph{conserved pre-symplectic form} over an arbitrary space-like surface $M$,
\begin{equation}\label{PreSymplecticForm}
\tilde{\Omega} (\delta_{1} , \delta_{2} ) := \int_{M} J (\delta_{1}, \delta_{2}) = - \frac{1}{ \kappa} \int_{M}  \delta_{2} e^{I} \wedge \delta_{1} \omega_{I} - \delta_{1} e^{I} \wedge \delta_{2} \omega_{I}
\end{equation}
Once we have $\tilde{\Omega} (\delta_{1} , \delta_{2} )$, we can analyse the symmetries of the theory and their associated conserved charges. In particular we are interested in the conserved charge associated with the asymptotic time translations, i.e. the ADM energy.

Since one of our goals is to compare the resulting expression for the energy through the covariant and canonical formalism, we need to be sure that the conventions in both schemes are in agreement. We discuss this point in the next part.

\subsubsection{Link between covariant and canonical approaches}

The symplectic structure is essential in order to have a Hamiltonian description. In a coordinate basis associated with the configuration variables, the fields $\phi^{A}$, the symplectic form can also be defined by
\begin{equation}
\bar{\Omega} := \mathrm{d} \Pi_{A} \wedge \mathrm{d} \phi^{A},
\end{equation}
where $\Pi_{A}$ is the momenta canonically conjugated to $\phi^{A}$. This $\bar{\Omega}$ is consistent with all our derivations in the covariant phase space. But, up to now, we have not specified `what our variables are', namely $\phi^{A}$ and $\Pi_{A}$. 

It is well known that in the first order formulation of general relativity one of our configuration variables is the canonically conjugated variable to the other. For instance, in the connection-dynamics approach, $\omega$ is chosen to be the configuration variable and, as it turns out,  $e$ happens to be its canonical momenta. The role of the variables is inverted if we choose the geometrodynamics picture. 

To compare with the results obtained by the canonical formalism, first we have to decide if we want to work in the connection or geometrodynamics approach. In this contribution we choose the former one, that is $\phi^{A} = \omega^{I}$ and $\Pi_{A} = e_{I}$. From (\ref{PreSymplecticForm}) we have then,
\begin{equation}\label{PreSymplecticForm-AgreesCanonical}
\tilde{\Omega} (\delta_{1} , \delta_{2} ) = - \frac{1}{2 \kappa} \int_{M}  \delta_{2} \underbrace{e^{I}}_{\Pi^{A}} \wedge \delta_{1} \underbrace{\omega_{I}}_{\phi^{A}} - \delta_{1} e^{I} \wedge \delta_{2} \omega_{I} = - \bar{\Omega}
\end{equation}
We conclude then that in order to compare our expressions for the energy,  we 
have to set $\bar{\Omega} = - \tilde{\Omega}$. From now on, this is the choice we shall make.

\subsection{The Hamiltonian and the energy}

Consider infinitesimal diffeomorphisms generated by a vector field $\xi$, these diffeomorphisms induce an infinitesimal change in the fields given by $\delta_{\xi} := ( \pounds_{\xi} e , \pounds_{\xi} \omega)$. 

We say that $\xi$ is a Hamiltonian vector field iff $\bar{\Omega} (\delta , \delta_{\xi})$ is closed, $\mathrm{d} \!\!\!\! \mathrm{d}\, \Omega = 0$, and the Hamiltonian $H_{\xi}$ is defined by,
\begin{equation}\label{CovariantHamiltonianDef}
\bar{\Omega} (\delta , \delta_{\xi}) = \delta H_{\xi} = \mathrm{d} \!\!\!\! \mathrm{d}\, H.
\end{equation}

Where $\mathrm{d} \!\!\!\! \mathrm{d}\,$ is the exterior derivative on the covariant phase space\footnote{see \cite{crv1,crv2} for further details and definitions.}, which is different from the exterior derivative on spacetime $\mathrm{d}$. 

So $H_{\xi}$ is a conserved quantity along the flow generated by $\xi$. We consider the case when $\xi$ generates asymptotic time translations of the space-time, which induces time evolution on the covariant phase space generated by the vector field $\delta_{\xi} := ( \pounds_{\xi} e , \pounds_{\xi} \omega)$. In this case, $H_{\xi}$ is the energy.

\subsubsection{The energy}

From eq. (\ref{PreSymplecticForm}) and (\ref{CovariantHamiltonianDef}),
\begin{eqnarray}
\bar{\Omega} (\delta , \delta_{\xi} ) &=& - \tilde{\Omega} (\delta , \delta_{\xi} ) = \frac{1}{ \kappa} \int_{M}  \delta_{\xi} e^{I} \wedge \delta \omega_{I} - \delta e^{I} \wedge \delta_{\xi} \omega_{I}\\
&=& \frac{1}{ \kappa} \int_{M}  \pounds_{\xi} e^{I} \wedge \delta \omega_{I} - \delta e^{I} \wedge \pounds_{\xi} \omega_{I}
\end{eqnarray}
by using $\pounds_{\xi} \phi^{A} = \xi \cdot \mathrm{d} \phi^{A} + \mathrm{d} (\xi \cdot \phi^{A}) $
\begin{equation}
\bar{\Omega} (\delta , \delta_{\xi}) = \frac{1}{ \kappa} \int_{M} \left[ (\xi \cdot \mathrm{d} e^{I}) \wedge \delta \omega_{I} + \mathrm{d}(\xi \cdot e^{I}) \wedge \delta \omega_{I}  - \delta e^{I} \wedge (\xi \cdot \mathrm{d} \omega_{I})  -  \delta e^{I} \wedge \mathrm{d} ( \xi \cdot \omega_{I})  \right].
\end{equation}
Now we have to use that at infinity $\xi$ should approach a time-translation Killing vector field of the asymptotic flat spacetime. In particular this means that in the asymptotic region $\xi ^{a} $ is orthogonal to the spacelike surface. Therefore $\xi \cdot e^{I} = e_{0} ^{I}$, $\xi  \cdot \omega_{I} = \omega_{0I}$ but for the leading term we have seen $\,^{1} \bar{\omega}_{0I} = 0$, also $\mathring{\bar{\mathcal{D}}}_{a} \, ^{0} e_{b} ^{I}$ only has spatial components so $\xi \cdot \mathrm{d} \,^{0} e^{I} = 0$. With this at hand we can see that\footnote{This is the only non-vanishing term to first order.},
\begin{eqnarray}\label{Presymplectic1}
\bar{\Omega} (\delta , \delta_{\xi}) &=& \frac{1}{ \kappa} \int_{M} \mathrm{d}(\xi \cdot e^{I}) \wedge \delta \omega_{I}\\
&=& \frac{1}{ \kappa}   \int_{M} \mathrm{d} \left[ (\xi \cdot e^{I}) \delta \omega_{I} \right] - (\xi \cdot e^{I}) \mathrm{d} \delta \omega_{I}
\end{eqnarray}
Note that the second term of the previous equation in components becomes,
\begin{equation}
- \int_{M}  \, ^{0} \bar{e}_{0} ^{I} \mathring{\bar{\mathcal{D}}}_{[\bar{b}|} \delta \omega_{|\bar{c}]I} \varepsilon^{\bar{b}\bar{c}} r \d r \d \theta 
\end{equation}
but
\begin{equation}
 \mathring{\bar{\mathcal{D}}}_{[\bar{b}|} \delta \omega_{|\bar{c}]} ^{M} =  \delta \beta \left[ r^{-2} \partial_{[\bar{b}| } r \partial_{\bar{a} } r + r^{-1} \partial_{[\bar{b}|} \partial_{\bar{a}} r  \right] \varepsilon_{LK}\,^{M} \,^{0} \bar{e}^{K} _{\bar{d}} \,^{0} \bar{e}^{L} _{|\bar{c}]} \eta^{\bar{a} \bar{d}} ,
\end{equation}
so
\begin{equation}
\, ^{0} \bar{e}_{0} ^{I} \mathring{\bar{\mathcal{D}}}_{[\bar{b}|} \delta \omega_{|\bar{c}]I} \varepsilon^{\bar{b}\bar{c}}  =   \delta \beta \left[ r^{-2} \partial_{[\bar{b}| } r \partial_{\bar{a} } r + r^{-1} \partial_{[\bar{b}|} \partial_{\bar{a}} r  \right] \underbrace{\varepsilon_{LKI} \,^{0} \bar{e}^{K} _{\bar{d}} \,^{0} \bar{e}^{L} _{|\bar{c}]} \, ^{0} \bar{e}_{0} ^{I}}_{\bar{e} \tilde{\varepsilon}_{|\bar{c}] \bar{d} 0 } } \eta^{\bar{a} \bar{d}} \underbrace{\varepsilon^{\bar{b}\bar{c}}}_{\frac{\tilde{\varepsilon}^{0 \bar{b}\bar{c}}}{r}}  ,
\end{equation}
here $\bar{e} = \sqrt{- \eta} = 1$ where $\eta_{ab}$ is the Minkowski metric associated with the fixed frame $\bar{e}_{I} ^{a}$ at the asymptotic region, also $\tilde{\varepsilon}_{|\bar{c}] \bar{d} 0 }\tilde{\varepsilon}^{0 \bar{b}\bar{c}} = -2 \delta_{|\bar{d}] } ^{\bar{b} } $. Thus by antisymmetry in the space-time indices this term vanishes. 

From (\ref{Presymplectic1}) and the previous argument the presymplectic form is,
\begin{eqnarray}\label{Omega-energy}
\nonumber \bar{\Omega} (\delta , \delta_{\xi}) &=& \frac{1}{ \kappa}  \int_{M} \mathrm{d} \left[ (\xi \cdot e^{I}) \delta \omega_{I} \right] =  \frac{1}{ \kappa} \int_{\partial M}  (\xi \cdot e^{I}) \delta \omega_{I} \\
&=& \lim_{r \rightarrow \infty} \left[ \frac{1}{\kappa} \int_{\partial M} \,^{0} e_{0} ^{I} \delta  \frac{ \,^{1} \omega_{\bar{c}I}}{r} \varepsilon^{0\bar{c}} r \d \theta + \int_{\partial M} {\cal{O}} (r^{-1}) \d \theta \right],
\end{eqnarray}
with
\begin{equation}
\delta \left( \frac{\,^{1} \omega_{\bar{c}} ^{M}}{r} \right) = \frac{1}{2r} \delta \beta  \partial_{\bar{a}} r \varepsilon_{LK} \, ^{M} \,^{0} \bar{e} ^{K} _{\bar{d}} \,^{0} \bar{e} ^{L} _{\bar{c}} \eta^{\bar{a} \bar{d}} .
\end{equation}

Then, by (\ref{CovariantHamiltonianDef}), the variation of the Hamiltonian, and therefore of its corresponding associated conserved quantity, the energy, is
\begin{eqnarray}\label{BigExpression-EnergyCovariant}
\nonumber \delta H_{\xi} = \bar{\Omega} (\delta , \delta_{\xi}) &=&  \frac{1}{2 \kappa} \lim_{r \rightarrow \infty} \int_{\partial M} \,^{0} \bar{e}_{0} ^{I} \left(  \delta \beta r^{-1} \partial_{\bar{a}} r \varepsilon_{LKI} \,^{0} \bar{e} ^{K} _{\bar{d}} \,^{0} \bar{e} ^{L} _{\bar{c}} \eta^{\bar{a} \bar{d}} \right) \varepsilon^{0\bar{c}} r \d \theta\\
&=&  \frac{1}{2 \kappa} \lim_{r \rightarrow \infty} \int_{\partial M} \frac{1}{r} \delta \beta  \underbrace{\left( \varepsilon_{LKI}  \,^{0} \bar{e}_{0} ^{I}  \,^{0} \bar{e} ^{K} _{\bar{d}} \,^{0} \bar{e} ^{L} _{\bar{c}} \right)}_{\bar{e} \tilde{\varepsilon}_{\bar{c} \bar{d} 0 } }  \eta^{\bar{a} \bar{d}}  \partial_{\bar{a}} r \varepsilon^{0\bar{c}} r \d \theta  
\end{eqnarray}
here we are using the identity $\varepsilon_{LKI}  \,^{0} \bar{e}_{0} ^{I}  \,^{0} \bar{e} ^{K} _{\bar{d}} \,^{0} \bar{e} ^{L} _{\bar{c}}  =  \bar{e} \tilde{\varepsilon}_{0 \bar{c} \bar{d}}$ where
$\,^{0} \bar{e} = \sqrt{- \eta} = 1$ with $\eta$ the determinant of $\eta_{ab}$, the Minkowski metric associated with the fixed frame $\bar{e}_{I} ^{a}$ at the asymptotic region. Taking into account the fall-off conditions on $e_{a} ^{I}$, its determinant, $e$, will decay as $e = \,^{0} \bar{e} + O(r^{-1})$, then $\bar{e} \tilde{\varepsilon}_{\bar{c} \bar{d} 0 } = \left[ e \varepsilon_{\bar{c} \bar{d} 0 } - \varepsilon_{\bar{c} \bar{d} 0 } O(r^{-1})  \right]$. Also $\partial^{\bar{a}} r =: r^{\bar{a}}$ can be seen as the normal to the cylinders $r = const$ and $\tilde{\varepsilon}_{abc} r^{c} = \tilde{\varepsilon}_{ab}$.
With all this we can see that the previous equation (\ref{BigExpression-EnergyCovariant}) is,
\begin{equation}
\delta H_{\xi} =  \frac{1}{2 \kappa} \lim_{r \rightarrow \infty} \int_{\partial M} \left[ \frac{1}{r} \delta \beta \varepsilon_{\bar{c} \bar{d} 0 }  r^{\bar{d} } \varepsilon^{0\bar{c}} r \d\theta  + O(r^{-1})  \right] =  \frac{1}{2 \kappa} \int_{\partial M}   \delta \beta   \underbrace{\varepsilon_{0 \bar{c} } \varepsilon^{ 0 \bar{c}}}_{1}  \d \theta =  \frac{1}{2 \kappa}  \delta \beta \int_{\partial M} \d \theta
\end{equation}
Also note that $\partial M = C_{t}$, $M$ a space like slice at ``time'' $t$, and $C_{t}$ a circle with radius $r$ at time $t$. We can write the expression for the energy,
\begin{equation}
\delta H_{\xi} = \frac{ \delta \beta}{2 \kappa} \int_{C_{t}} \d \theta
\end{equation}
taking $\kappa = 8 \pi G$,
\begin{equation}
\delta H_{\xi} = \frac{ \delta \beta}{2 (8 \pi G)} 2 \pi =  \frac{\delta \beta}{8 G}.
\end{equation}
Since the previous expression only gives the variation, the energy will always be determined up to a constant, 
\begin{equation}\label{Energy-covariant-const}
E = \frac{ \beta}{16G} + \mathrm{const}', .
\end{equation}
Let us summarize the situation. By employing the covariant Hamiltonian formalism, we have reached an expression for the gradient of the Hamiltonian function on the covariant phase space, responsible for the Hamiltonian flow that generates asymptotic unit time translations. As is usually the case with the Hamiltonian formalism, this function is determined up to a constant. Here we are faced with several
choices. We could, for instance, follow \cite{Ashtekar-Varadarajan1994} and declare that Minkowski spacetime should have a vanishing energy. Since $\beta \in [0,2)$, we should then choose this constant to be zero for the energy of Minkowski space-time to vanish,
\begin{equation}
E \in \left[ 0,  \frac{1}{4G} \right].
\end{equation}
Although the CHF is elegant, it only provide us with the variation of the energy, so we have an 
indeterminacy in the election of the constant that may shift the region in which the energy is bounded. 
Of course, we are in principle allowed to make any other choice for the up to now arbitrary constant, 
unless we take some input that helps us select it.
That is why we shall analyse this action through the canonical 2+1 formalism, where the Hamiltonian 
{\emph is} completely determined by the Legendre transform. This is the subject of the following section.


\section{Canonical analysis}
\label{sec:5}

In the case of theories that can be formulated without the need of a metric, we have two choices for a $2+1$ decomposition. The first one, that we shall refer to as the \emph{Witten approach}\footnote{Following the nomenclature of \cite{Barbero-Varadarajan1994} referring to Witten's paper \cite{Witten1988}. For more details on the analysis in the case where there is no boundary see \cite{romano}.}, it does not need the existence of a metric. We only ask the spacetime $\mathcal{M}$ to be topologically $\Sigma \times R$ and that there exists a function $t$ (with nowhere vanishing gradient $(\d t)_{a}$) such that each $t$= const surface $M_{t}$ is diffeomorphic to $\Sigma$. Also, one assumes the existence a flow defined by a vector field $t^{a}$ satisfying $t^{a} (\d t)_{a} = 1$, which allow us  to define ``evolution'', although $t$ does not necessarily have the interpretation of time\footnote{Since the $2+1$ Palatini action based on an arbitrary Lie group $G$ (\ref{SPB}) is a theory independent of a spacetime metric, we can still define evolution from one $t=const$ surface to the next using the Lie derivative along $t^{a}$.}.

The second approach, that we shall refer to as the \emph{Ashtekar-Barbero-Varadarajan} approach\footnote{In \cite{Barbero-Varadarajan1994} the authors discuss the differences in the canonical analysis, particularly in the constraints, following Witten's vs Ashtekar's approaches. That is why we call it Ashtekar-Barbero-Varadarajan approach.} follows closely the $3+1$ decomposition of the first order variables. In it, besides the elements of the Witten approach, we are also assuming the existence of a metric $g_{ab}$ and therefore a unit normal $n^a$ to the Cauchy surfaces. 
This introduces additional information to that in Witten's decomposition. In particular, we can decompose any tensor into its normal and tangential part, and in particular $t^{a}$ can be decomposed as $t^{a} = Nn^{a} + N^{a}$, where $N$ and $N^{a}$ are the lapse and shift functions. Now we have additional information, namely the freedom of choosing any foliation and any vector field $t^{a}$, that is coded in the lapse and shift functions.

A comment on notation is in order. In what follows we use $\tilde{\varepsilon}^{abc}$ as the Levi-Civita tensor density of weight $+1$ instead of  $\tilde{\eta}^{abc}$, more commonly used in the $3-$ dimensional case, this to avoid confusion with the flat metric $\bar{\eta}_{ab}$ (\ref{eta}), or with the Minkowki metric (either with internal or spacetime indices). 
When we write $\tilde{\varepsilon}^{abc}$ in the action we assume it is accompanied with its respective $d^{3}x$, but we do not write it in order to simplify notation. Only when dealing with the Levi-Civita tensor, $\varepsilon^{abc}$, related with the tensor density by $\tilde{\varepsilon}^{abc} = (s) \sqrt{|g|} \varepsilon^{abc}$ (with $g$ the determinant of the spacetime metric and $s$ the signature of the metric), we write the volume element explicitly. The same convention will be used for the $\tilde{\varepsilon}^{ab}$ and $\tilde{\varepsilon}^{a}$. Finally, 
 we shall refer to a Cauchy slice as $M$ following the notation in \cite{crv1,crv2}.

\subsection{Witten's approach}
\label{Witten-approach}

In order to make the canonical analysis (a la Witten) of the 3-dimensional Palatini action, we write the action (\ref{LIP}) it in components,
 \begin{eqnarray}\label{Palatini-arbitraryG}
 S_{PB} [e, \omega ] &=& - \frac{1}{2 \kappa} \int_{\mathcal{M}} \tilde{\varepsilon}^{abc} e_{aI} F_{bc} ^{I} -  \frac{1}{ \kappa} \int_{\partial \mathcal{M} } e_{aI} \omega_{b} ^{I} \tilde{\varepsilon}^{ab} - \frac{\alpha}{\kappa} \int_{\partial \mathcal{M} } \frac{1}{ n \cdot n} \varepsilon^{IKL} e_{aI} n_{K} \mathring{\bar{\mathcal{D}}} _{b} n_{L} \tilde{\varepsilon}^{ab}\\
 &=& - \frac{1}{2 \kappa} \int_{\mathcal{M}} \tilde{\varepsilon}^{abc} e_{aI} F_{bc} ^{I} -  \frac{1}{ \kappa} \int_{\partial \mathcal{M} } e_{aI} \omega_{b} ^{I} \tilde{\varepsilon}^{ab} + \frac{\alpha}{\kappa} \int_{\partial \mathcal{M} } \frac{1}{\sqrt{ n \cdot n}} \varepsilon^{IL} e_{aI} \mathring{\bar{\mathcal{D}}} _{b} n_{L} \tilde{\varepsilon}^{ab}
 \end{eqnarray}
For this decomposition we shall follow the analysis in \cite{romano}, taking enough care of the boundary term, the one coming from the Palatini action and the boundary terms in (\ref{Palatini-arbitraryG}). Using that $\tilde{\varepsilon}^{abc} = 3 t^{[a} \tilde{\varepsilon}^{bc]} \d t$ and $\tilde{\varepsilon}^{ab} = 2 t^{[a} \tilde{\varepsilon}^{b]} \d t$
\begin{eqnarray}
\nonumber S_{PB} [e, \omega ] &=& -   \frac{1}{2 \kappa}\int \d t \int_{M} ( t^{a} \tilde{\varepsilon}^{bc} + t^{b} \tilde{\varepsilon}^{ca}  + t^{c} \tilde{\varepsilon}^{ab}) e_{aI} F_{bc}^{I} - \frac{1}{ \kappa} \int \d t \int_{C_{t}} (t^{a} \tilde{\varepsilon}^{b} - t^{b} \tilde{\varepsilon}^{a}) e_{aI} \omega_{b} ^{I} \\
&&+ \frac{\alpha}{ \kappa} \int \d t \int_{C_{t}} (t^{a} \tilde{\varepsilon}^{b} - t^{b} \tilde{\varepsilon}^{a}) \frac{1}{\sqrt{ n \cdot n}} \varepsilon^{IL} e_{aI} \mathring{\bar{\mathcal{D}}} _{b} n_{L}  \\
\nonumber &=& - \frac{1}{ \kappa}\int \d t \int_{M} \left[  \frac{1}{2} \underbrace{(t^{a} e_{aI})}_{(t \cdot e)^{I}} F_{bc}^{I} \tilde{\varepsilon}^{bc} + t^{b}\tilde{\varepsilon}^{ca} e_{aI} F_{bc} ^{I}   \right] \\
&&- \frac{1}{\kappa} \int \d t \int_{C_{t}} \left[  \underbrace{ (t^{a} e_{aI})}_{(t \cdot e)^{I}} \omega_{b} ^{I} \tilde{\varepsilon}^{b} -  \underbrace{(t^{b} \omega_{b} ^{I})}_{(t\cdot \omega)^{I}} e_{aI} \tilde{\varepsilon}^{a}   \right]\\
&&+ \frac{\alpha}{ \kappa} \int \d t \int_{C_{t}}  \frac{1}{\sqrt{ n \cdot n}} \varepsilon^{IL} \left[( t^{a}  e_{aI} )\mathring{\bar{\mathcal{D}}} _{b} n_{L} \tilde{\varepsilon}^{b} -    (t^{b} \mathring{\bar{\mathcal{D}}} _{b} n_{L}) e_{aI} \tilde{\varepsilon}^{a} \right]
\end{eqnarray}
Taking into account the following standard relations,
\begin{eqnarray}
F_{bc} ^{I} &=& 2 \partial_{[b} \omega_{c]} ^{I} + [\omega_{b} , \omega_{c}]^{I} = \partial_{b} \omega_{c} - \partial_{c} \omega_{b} + [\omega_{b} , \omega_{c}]^{I}\\
\mathcal{D}_{b} \omega_{c} ^{I} &=& \partial_{b} \omega_{c} ^{I}  + [\omega_{b} , \omega_{c}]^{I}\\
t^{b} F_{bc} ^{I} &=& \pounds_{\vec{t}} \omega_{c} ^{I} - \mathcal{D}_{c} (t\cdot \omega)^{I} 
\label{tF-igual-LieD-menos-CovD}
\end{eqnarray}
the second term of the bulk part can be written as,
\begin{eqnarray}
\tilde{\varepsilon}^{ca} e_{aI} t^{b} F_{bc} ^{I}  &=& (\pounds_{\vec{t}} \omega_{c} ^{I}) \tilde{\varepsilon}^{ca} e_{aI} - \mathcal{D}_{c} (\omega \cdot t)^{I} \tilde{\varepsilon}^{ca} e_{aI}\\
&=& (\pounds_{\vec{t}} \omega_{c} ^{I}) \tilde{\varepsilon}^{ca} e_{aI} - \mathcal{D}_{c} [(\omega \cdot t)^{I} \tilde{\varepsilon}^{ca} e_{aI}^{I}] + (\omega \cdot t)^{I} \mathcal{D}_{c} (\tilde{\varepsilon}^{ca} e_{aI}).
\end{eqnarray}
Then the action takes the form,
\begin{eqnarray}
\nonumber S_{PB} [e, \omega ] &=& -   \frac{1}{ \kappa}\int \d t \int_{M} \left[  \frac{1}{2} \underbrace{(t^{a} e_{aI})}_{(t \cdot e)_{I}} F_{bc}^{I} \tilde{\varepsilon}^{bc} + (\pounds_{\vec{t}} \omega_{c} ^{I}) \tilde{\varepsilon}^{ca} e_{aI} - \mathcal{D}_{c} [(\omega \cdot t)^{I} \tilde{\varepsilon}^{ca} e_{aI}] + (\omega \cdot t)^{I} \mathcal{D}_{c} (\tilde{\varepsilon}^{ca} e_{aI}) \right] \\
&&- \frac{1}{\kappa} \int \d t \int_{C_{t}} \left[  \underbrace{ (t^{a} e_{aI})}_{(t \cdot e)_{I}} \omega_{b} ^{I} \tilde{\varepsilon}^{b} - \underbrace{(t^{b} \omega_{b} ^{I})}_{(t\cdot \omega)^{I}} e_{aI} \tilde{\varepsilon}^{a}   \right]\\
&&+ \frac{\alpha}{ \kappa} \int \d t \int_{C_{t}}  \frac{1}{\sqrt{ n \cdot n}} \varepsilon^{IL} \left[( t^{a}  e_{aI} )\mathring{\bar{\mathcal{D}}} _{b} n_{L} \tilde{\varepsilon}^{b} -    (t^{b} \mathring{\bar{\mathcal{D}}} _{b} n_{L}) e_{aI} \tilde{\varepsilon}^{a} \right]\\
\nonumber &=&  - \frac{1}{ \kappa}\int \d t \int_{M} \left[  \frac{1}{2} \underbrace{(t^{a} e_{aI})}_{(t \cdot e)_{I}} F_{bc}^{I} \tilde{\varepsilon}^{bc} + (\pounds_{\vec{t}} \omega_{c} ^{I}) \tilde{\varepsilon}^{ca} e_{aI}  + (\omega \cdot t)^{I} \mathcal{D}_{c} (\tilde{\varepsilon}^{ca} e_{aI}) \right] \\
&& + \frac{1}{ \kappa}\int \d t \int_{M} \mathcal{D}_{c} [(\omega \cdot t)^{I} \tilde{\varepsilon}^{ca} e_{aI}] - \frac{1}{\kappa} \int \d t \int_{C_{t}} \left[  \underbrace{ (t^{a} e_{aI})}_{(t \cdot e)_{I}} \omega_{b} ^{I} \tilde{\varepsilon}^{b} -  \underbrace{(t^{b} \omega_{b} ^{I})}_{(t\cdot \omega)^{I}} e_{aI} \tilde{\varepsilon}^{a}   \right]\\
&&+ \frac{\alpha}{ \kappa} \int \d t \int_{C_{t}}  \frac{1}{\sqrt{ n \cdot n}} \varepsilon^{IL} \left[( t^{a}  e_{aI} )\mathring{\bar{\mathcal{D}}} _{b} n_{L} \tilde{\varepsilon}^{b} -    (t^{b} \mathring{\bar{\mathcal{D}}} _{b} n_{L}) e_{aI} \tilde{\varepsilon}^{a} \right].
\end{eqnarray}

Strictly speaking we begin with an action valid for \emph{any} Lie group ($e$ is not related to the metric unless we identify the group with $SO(2,1)$ so this action can be defined without the need of a metric), since in Witten's decomposition we are not assuming the existence of a metric.  

In order to proceed with the Legendre transformation we need to calculate the momenta,
\begin{equation}
\Pi_{I} ^{c} = \frac{\delta \mathcal{L} }{\delta (\pounds_{\vec{t}} \omega_{c} ^{I}  )  } = \frac{1}{\kappa} \tilde{\varepsilon} ^{ca} e_{aI},
\end{equation}
then the canonical Hamiltonian is\footnote{Note that the bulk part of this Hamiltonian coincides with that given in \cite{romano}.},
\begin{eqnarray}\label{Hamiltonian-Witten}
\nonumber H[e, \omega] &=& \int_{M} \left[ (\pounds_{\vec{t}} \omega_{c} ^{I}  )\Pi_{I} ^{c}  - \mathcal{L} \right]\\
\nonumber &=& + \frac{1}{ \kappa} \int_{M} \left[  \frac{1}{2} \underbrace{(t^{a} e_{aI})}_{(t \cdot e)_{I}} F_{bc}^{I} \tilde{\varepsilon}^{bc} + (\omega \cdot t)^{I} \mathcal{D}_{c} (\tilde{\varepsilon}^{ca} e_{aI}) \right] \\
\nonumber && - \frac{1}{ \kappa} \int_{M} \mathcal{D}_{c} [(\omega \cdot t)^{I} \tilde{\varepsilon}^{ca} e_{aI}] + \frac{1}{\kappa}  \int_{C_{t}} \left[  \underbrace{ (t^{a} e_{aI})}_{(t \cdot e)_{I}} \omega_{b} ^{I} \tilde{\varepsilon}^{b} -  \underbrace{(t^{b} \omega_{b} ^{I})}_{(t\cdot \omega)^{I}} e_{aI} \tilde{\varepsilon}^{a}   \right]\\
&&- \frac{\alpha}{ \kappa}  \int_{C_{t}}  \frac{1}{\sqrt{ n \cdot n}} \varepsilon^{IL} \left[( t^{a}  e_{aI} )\mathring{\bar{\mathcal{D}}} _{b} n_{L} \tilde{\varepsilon}^{b} -    (t^{b} \mathring{\bar{\mathcal{D}}} _{b} n_{L}) e_{aI} \tilde{\varepsilon}^{a} \right].\end{eqnarray}
We can see that the following constraints
\begin{equation}\label{Witten-constraints}
F_{bc}^{I} \tilde{\varepsilon}^{bc} \approx 0 \,\,\,\,\, \mathrm{and} \,\,\,\,\,  \mathcal{D}_{c} (\tilde{\varepsilon}^{ca} e_{aI}) \approx 0,
\end{equation}
are first class, and also they are the pull-back to $M$ with $\tilde{\varepsilon}^{ab}$ of the equations of motion (\ref{PalatiniEoM}).

On the constraint surface,
\begin{eqnarray}
\nonumber H [e, \omega] &=& - \frac{1}{ \kappa} \int_{M} \mathcal{D}_{c} [(\omega \cdot t)^{I} \tilde{\varepsilon}^{ca} e_{aI}] + \frac{1}{\kappa}  \int_{C_{t}} \left[  \underbrace{ (t^{a} e_{aI})}_{(t \cdot e)_{I}} \omega_{b} ^{I} \tilde{\varepsilon}^{b} -  \underbrace{(t^{b} \omega_{b} ^{I})}_{(t\cdot \omega)^{I}} e_{aI} \tilde{\varepsilon}^{a}   \right]\\
&&- \frac{\alpha}{ \kappa}  \int_{C_{t}}  \varepsilon^{IL} \left[( t^{a}  e_{aI} )\mathring{\bar{\mathcal{D}}} _{b} \frac{ n_{L}}{\sqrt{ n \cdot n}} \tilde{\varepsilon}^{b} -    \left(t^{b} \mathring{\bar{\mathcal{D}}} _{b} \frac{ n_{L}}{\sqrt{ n \cdot n}}\right) e_{aI} \tilde{\varepsilon}^{a} \right].
\end{eqnarray}
that is, the boundary terms are the only non-vanishing terms.

Now if we take into account the asymptotically flat boundary conditions, the leading term of $(\omega \cdot t)^{I} $ is zero and also $t^{b} \mathring{\bar{\mathcal{D}}} _{b} (r^{c} \, ^{0} e_{cL} ) = 0$. In the timelike boundary as well as in the boundary of $M$ (circles for each time $t$, $C_{t}$) the normal to the surface is $r^{a}$, then $n_{L} / \sqrt{n \cdot n } = r^{c} e_{cL}$. So the only non-vanishing leading term comes from,
\begin{equation}\label{H-nonvanishingleadingtermscomefrom}
H [e, \omega] =   \frac{1}{ \kappa}  \int_{C_{t}}  (t \cdot e)_{I} \omega_{b} ^{I} \tilde{\varepsilon}^{b} - \frac{\alpha}{ \kappa}  \int_{C_{t}}   \varepsilon^{IL} ( t^{a}  e_{aI} ) \underbrace{\mathring{\bar{\mathcal{D}}} _{b} (r^{c} e_{cL} )}_{ r^{c} \mathring{\bar{\mathcal{D}}} _{b} e_{cL} +  e_{cL} \mathring{\bar{\mathcal{D}}} _{b} r^{c} } \tilde{\varepsilon}^{b} .
\end{equation}

As in the covariant case, if we want this Hamiltonian to generate asymptotic time translations and therefore its conserved quantity to be the \emph{energy}, $t^{a}$ has to approach a unit time-translation Killing vector field of the asymptotic flat spacetime, which also translates into $t$ being orthogonal to $M$. Using this and the fall-off conditions (\ref{AFfalloff-cotriad}) and (\ref{FalloffOmegaZero}), the Hamiltonian is given by\footnote{The term (that comes from eq. (\ref{H-nonvanishingleadingtermscomefrom})),
\begin{eqnarray}
\nonumber \, ^{Leading} \lim_{r \rightarrow \infty} - \frac{\alpha}{ \kappa}  \int_{C_{t}}  \varepsilon^{IL} ( t^{a}  e_{aI} ) r^{c} \mathring{\bar{\mathcal{D}}} _{b} e_{cL} \tilde{\varepsilon}^{b}  &=& \lim_{r \rightarrow \infty} \left[ - \frac{\alpha}{ \kappa}  \int_{C_{t}}  \varepsilon^{IL} \, ^{0} e_{0I}  r^{c} ( - \frac{\beta}{2r} r^{ - \beta /2 } \partial_{b} r \, ^{0} \bar{e}_{\bar{c}L} \delta^{\bar{c}} _{c} )\tilde{\varepsilon}^{0b} + {\cal O}(r^{- 1})\right] \\
\nonumber &=& \lim_{r \rightarrow \infty}\left[   \frac{\alpha \beta}{ 2\kappa}  \int_{C_{t}}   \underbrace{\varepsilon^{IL} \, ^{0} e_{0I} \, ^{0} \bar{e}_{\bar{c}L}}_{\bar{e} \tilde{\varepsilon}_{0\bar{c} }}   \frac{1}{r} r^{ - \beta /2 } r^{\bar{c}}  \partial_{b} r    \tilde{\varepsilon}^{0b}  + {\cal O}(r^{- 1})\right] \\
\nonumber &=& \lim_{r \rightarrow \infty} \left[   \frac{\alpha \beta}{ 2\kappa}  \int_{C_{t}}     \frac{1}{r} r^{ - \beta /2 } r^{\bar{c}}  \partial_{b} r \underbrace{ \tilde{\varepsilon}_{0\bar{c} } \tilde{\varepsilon}^{0b} }_{\delta_{\bar{c}} ^{b}}   r \d\theta  + {\cal O}(r^{- 1})\right] \\
\nonumber &=& \lim_{r \rightarrow \infty} \left[   \frac{\alpha \beta}{ 2\kappa}  \int_{C_{t}}      r^{ - \beta /2 } (+1)   \d\theta + {\cal O}(r^{- 1})\right] \\
\nonumber &=& \lim_{r \rightarrow \infty} \left[   \frac{\alpha \beta}{ 2\kappa} r^{ - \beta /2 } 2\pi + {\cal O}(r^{- 1})\right]  \\
\nonumber &=& \lim_{r \rightarrow \infty}  \left[  {\cal O}(r^{- \beta / 2})  + {\cal O}(r^{- 1})\right]  = 0   \,\,\,\,\, iff\,\,\,\,\, \beta > 0\\
\end{eqnarray}
},
\begin{equation}\label{H-Wittendecomposition}
H [e, \omega] =   \lim_{r \rightarrow \infty} \left[  \underbrace{  \int_{C_{t}} \frac{1}{ \kappa} \,^{0} e_{0I} \frac{\,^{1} \omega_{\bar{b}} ^{I}}{r} \tilde{\varepsilon}^{\bar{b}} }_{H_{1}} \underbrace{- \frac{\alpha}{ \kappa}  \int_{C_{t}}  \varepsilon^{IL}  \, ^{0}  e_{0I}  \left( \, ^{0} e_{cL} \mathring{\bar{\mathcal{D}}} _{b} r^{c} \right) \tilde{\varepsilon}^{b} }_{H_{2}} + {\cal O}(r^{- \beta / 2}) \right]
\end{equation}
For the first term of the right hand side of previous equation, since the volume element associated to $C_{t}$ goes as $r \d\theta$, the leading term of the previous equation does not depend on $r$, and the next to leading terms go as ${\cal O}(r^{-1})$ so in the limit they vanish leaving us with just the leading term,
\begin{equation}\label{H1-Wittens}
H_{1} =  \frac{1}{ \kappa} \lim_{r \rightarrow \infty}  \int_{C_{t}}  \,^{0} e_{0I} \frac{\,^{1} \bar{\omega}_{\bar{b}} ^{I}}{r} \tilde{\varepsilon}^{\bar{b}} =  \frac{1}{2 \kappa} \lim_{r \rightarrow \infty}  \int_{C_{t}}  \,^{0} e_{0I} \frac{1}{r}  \beta  \partial_{\bar{a}} r \varepsilon_{L}\, ^{KI} \,^{0} \bar{e}_{K} ^{\bar{a}} \,^{0} \bar{e}_{\bar{b}} ^{L}  \tilde{\varepsilon}^{0\bar{b}}
\end{equation}
Note that, apart from $\delta \beta \leftrightarrow \beta$, this expression \emph{is the same} as (\ref{BigExpression-EnergyCovariant}). Using the same steps we can see that (taking $\kappa = 8 \pi G$),
\begin{equation}
H_{1} = \frac{ \beta}{2 \kappa} \int_{C_{t}} \d \theta = \frac{ \beta}{2 (8 \pi G)} 2 \pi =  \frac{ \beta}{8 G}.
\end{equation}

For the second term of the right hand side,
\begin{eqnarray}\label{H2-Wittens}
\nonumber H_{2} &=&  \lim_{r \rightarrow \infty}  \left[ - \frac{\alpha}{ \kappa}  \int_{C_{t}}   \varepsilon^{IL}  \, ^{0}  e_{0I}  \left( \, ^{0} e_{cL} \mathring{\bar{\mathcal{D}}} _{b} r^{c} \right) \tilde{\varepsilon}^{0b} \right]\\
\nonumber &=& - \frac{\alpha}{ \kappa} \lim_{r \rightarrow \infty}  \int_{C_{t}}   \underbrace{ \varepsilon^{IL}  \, ^{0}  e_{0I}  \, ^{0} e_{cL} }_{\bar{e}  \tilde{\varepsilon}_{0c} } \underbrace{\mathring{\bar{\mathcal{D}}} _{b} r^{c}}_{\partial_{b} r^{c}}  \tilde{\varepsilon}^{0b} \\
\nonumber &=& - \frac{\alpha}{ \kappa} \lim_{r \rightarrow \infty}  \int_{C_{t}}  \underbrace{ \tilde{\varepsilon}^{0b} \tilde{\varepsilon}_{0c}}_{\delta^{b} _{c}} (\partial_{b} r^{c} ) r \mathrm{d} \theta\\
&=& - \frac{\alpha}{ \kappa} \lim_{r \rightarrow \infty}  \int_{C_{t}} \underbrace{ (\partial_{c} r^{c} )}_{1/r} r \mathrm{d} \theta =  - \frac{\alpha}{2 \kappa}  \int_{C_{t}} 2 \d \theta 
\end{eqnarray}

Using (\ref{H1-Wittens}) and  (\ref{H2-Wittens}), we can see that the Hamiltonian (\ref{H-Wittendecomposition}) is given by,
\begin{equation}\label{Witten-lastHamiltonian}
H = H_{1} + H_{2} = \frac{ \beta}{2 \kappa} \int_{C_{t}} \d\theta - \frac{\alpha}{2 \kappa}   \int_{C_{t}} 2 \d\theta = - \frac{ 1}{2 \kappa} \int_{C_{t}} (2 \alpha - \beta) \d\theta.
\end{equation}
Let us summarize the situation. We have performed the 2+1 decomposition, {\emph a la} Witten, of the two 
actions we considered in Sec.~\ref{sec:3}. After performing the Legendre transform, the Hamiltonian is 
given by the boundary term of Eq.~(\ref{Witten-lastHamiltonian}). Recall that we have introduced a 
`switch' $\alpha$, that selects between the totally Lorentz invariant action ($\alpha=1$) and the 
generalized Palatini action ($\alpha=0$). 
The first obvious observation is that the Hamiltonian (and energy), depends 
on $\alpha$ and therefore, on the action we started with. Let us now analyse both cases.

Let us first consider the case when $\alpha = 1$, and note that we recover the results 
of \cite{Marolf-Patino2006},
\begin{equation}
H = - \frac{ 1}{2 \kappa} \int_{C_{t}} (2  - \beta) \d\theta.
\end{equation}
Following \cite{Ashtekar-Varadarajan1994, Marolf-Patino2006}, the parameter $\beta$ lies in the interval $\beta \in [0,2)$. From here we can conclude that the energy 
\be
E=\frac{1}{8G}(\beta - 2)\, ,
\label{energy-canonical}
\ee
is bounded from below and above, and lies within the interval,
$E \in \left[ -\frac{1}{4G}, 0 \right]$.
That is, all gravitational configurations have a negative energy, and in particular, Minkowski spacetime has an energy equal to $E_{\textrm{Mink}}= - 1/(4G)$.

The other case, namely when $\alpha = 0$, yields an energy $E_{\alpha=0}=\beta/(8G)$, that is always 
positive, with a zero value for the lower bound corresponding to Minkowski spacetime. 
In this sense one can observe that the energy found in the 
canonical description realizes the choice made by the authors of  \cite{Ashtekar-Varadarajan1994}.
This is the main result of this article. Let us now end this section with a few remarks.

\begin{enumerate}

\item Note that in both our analysis and in the one given in \cite{Marolf-Patino2006}, the starting point is a well posed action; the Palatini action with boundary term and the Einstein-Hilbert action with Gibbons-Hawking term respectively. Also, note that the addition of the
boundary term (\ref{AdditionalHarmlessBoundaryTerm}) is essential, within the first order action,
to be equivalent to the Einstein-Hilbert action with Gibbons-Hawking term. It is then not surprising that the LIP action leads to the same expression for the energy as in \cite{Marolf-Patino2006}.

\item Even though both actions, SPB and LIP, lead to the same classical equations of motion, the Einstein equations of motion, they do not completely agree at the Hamiltonian level, since they differ by a constant. 

\item It is important emphasize the difference between our result, where the Hamiltonian and therefore the energy is \emph{completely} determined by the Legendre transform, in contrast with the covariant  formalism where one only gets the variation of the Hamiltonian function, so the energy is only determined up to an additive constant (\ref{Energy-covariant-const}).

\end{enumerate}

In the next part we shall perform a different 2+1 splitting, that follows the standard decomposition and resembles the 3+1 case.

\subsection{Barbero-Varadarajan's approach}

As was the case in  Witten's decomposition, we shall begin with the well posed manifestly Lorentz invariant Palatini action,
\begin{equation}
S_{\textrm{LIP}} [e, \omega ] = - \frac{1}{2 \kappa} \int_{\mathcal{M}} \tilde{\varepsilon}^{abc} e_{aI} F_{bc} ^{I} -  \frac{1}{ \kappa} \int_{\partial \mathcal{M} } e_{aI} \omega_{b} ^{I} \tilde{\varepsilon}^{ab} - \frac{\alpha}{\kappa} \int_{\partial \mathcal{M} } \frac{1}{ n \cdot n} \varepsilon^{IKL} e_{aI} n_{K} \mathring{\bar{\mathcal{D}}} _{b} n_{L} \tilde{\varepsilon}^{ab}
\end{equation}

Using $\tilde{\varepsilon}^{abc} \varepsilon_{IJK} e_{c} ^{K} = 2e e_{I} ^{[a} e_{J} ^{b]}$, which implies $e \, \varepsilon^{LKM} e^{a} _{L} e^{b} _{K} =  \tilde{\varepsilon}^{abc} e_{c} ^{M}$. The well posed Palatini action can be written,
 \begin{equation}
 S_{\textrm{LIP}} [e, \omega ] = -  \frac{1}{2 \kappa} \int_{\mathcal{M}} e \varepsilon^{LKI} e_{L} ^{b} e_{K} ^{c} F_{bcI}  -  \frac{1}{ \kappa} \int_{\partial \mathcal{M} } e_{aI} \omega_{b} ^{I} \tilde{\varepsilon}^{ab} - \frac{\alpha}{\kappa} \int_{\partial \mathcal{M} } \frac{1}{ n \cdot n} \varepsilon^{IKL} e_{aI} n_{K} \mathring{\bar{\mathcal{D}}} _{b} n_{L} \tilde{\varepsilon}^{ab}
  \end{equation}
As we already mentioned, to make a standard $2+1$ \emph{decomposition}, we assume the existence of a metric and thus we can introduce a projector $q_{a} ^{b} = \delta_{a} ^{b} + n_{a} n^{b}$ which projects down all the fields in their spacelike and normal components respectively. In particular we can decompose $t^{a} = n^{a} N + N^{a}$. 

To begin with, we have to use $q_{a} ^{b}$ to project all the dynamical variables appearing in the action. First we shall decompose the integrand of the bulk term of the previous equation,
\begin{equation}
e \varepsilon^{LKI} e_{L} ^{b} e_{K} ^{c} F_{bcI} = e \varepsilon^{LKI} e_{L} ^{a} e_{K} ^{d} \delta_{a} ^{b} \delta_{d} ^{c} F_{bcI} = e \varepsilon^{LKI} e_{L} ^{a} e_{K} ^{d} (q_{a} ^{b} - n_{a} n^{b})( q_{d} ^{c} - n_{d}n^{c} )F_{bcI}
\end{equation}
with $q_{ab}$ the induced metric and $n^{a}$ the normal to the $2-$dimensional Cauchy slices. Now using $n^{a} = (t^{a} - N^{a})/N$,  also $\mathcal{E}_{a} ^{I} = q_{a} ^{b} e_{b} ^{I}$ and $\mathcal{F}_{ab} ^{I} = q_{a} ^{c} q_{b} ^{d} F_{cd} ^{I}$ are the projections of $e$ and $F$ to the Cauchy slice, and $n_{K} := n^{a} e_{aK}$, then the integrand of the bulk term becomes,
\begin{equation}
e \varepsilon^{LKI} e_{L} ^{b} e_{K} ^{c} F_{bcI} = e \varepsilon^{LKI} \left[ \mathcal{E}_{L} ^{b} \mathcal{E}_{K} ^{c} \mathcal{F}_{bcI} - \frac{2}{N} \mathcal{E}_{L} ^{b} n_{K} t^{c} F_{bcI} + \frac{2}{N} \mathcal{E}_{L} ^{b} n_{K} N^{c} \mathcal{F}_{bcI} \right].
\end{equation}
which implies that the decomposed bulk term is,
\begin{equation}\label{ABV-Bulktermdecomposed}
-  \frac{1}{2 \kappa} \int_{\mathcal{M}} e \varepsilon^{LKI} e_{L} ^{b} e_{K} ^{c} F_{bcI} = -  \frac{1}{2 \kappa} \int_{\mathcal{M}} e \varepsilon^{LKI} \left[ \mathcal{E}_{L} ^{b} \mathcal{E}_{K} ^{c} \mathcal{F}_{bcI} - \frac{2}{N} \mathcal{E}_{L} ^{b} n_{K} t^{c} F_{bcI} + \frac{2}{N} \mathcal{E}_{L} ^{b} n_{K} N^{c} \mathcal{F}_{bcI} \right].
\end{equation}

Now we shall decompose the boundary term,
\begin{equation}
 -  \frac{1}{ \kappa} \int_{\partial \mathcal{M} } e_{aI} \omega_{b} ^{I} \tilde{\varepsilon}^{ab} - \frac{\alpha}{\kappa} \int_{\partial \mathcal{M} } \frac{1}{ n \cdot n} \varepsilon^{IKL} e_{aI} n_{K} \mathring{\bar{\mathcal{D}}} _{b} n_{L} \tilde{\varepsilon}^{ab}.
\end{equation}
We begin with the integrand of the standard boundary term, $e_{aI} \omega_{b} ^{I} \tilde{\varepsilon}^{ab}$,
\begin{equation}
e_{aI} \omega_{b} ^{I} \tilde{\varepsilon}^{ab} = \delta_{a} ^{c} \delta_{b} ^{d} \tilde{\varepsilon}^{ab} e_{cI} \omega_{d} ^{I} = (q_{a} ^{c} - n_{a} n^{c})( q_{b} ^{d} - n_{b} n^{d}) e_{cI} \omega_{d} ^{I} \tilde{\varepsilon}^{ab}
\end{equation}
but $\tilde{\varepsilon}^{ab} = 2N n^{[a} \tilde{\varepsilon}^{b]} \d t$, then
\begin{eqnarray}
\nonumber e_{aI} \omega_{b} ^{I} \tilde{\varepsilon}^{ab} &=& N \left[q_{a} ^{c} q_{b} ^{d} e_{cI} \omega_{d} ^{I} (n^{a} \tilde{\varepsilon}^{b} - n^{b} \tilde{\varepsilon}^{a}) - q_{a} ^{c} n_{b} n^{d} e_{cI} \omega_{d} ^{I} (n^{a} \tilde{\varepsilon}^{b} - n^{b} \tilde{\varepsilon}^{a}) \right. \\
&& \left. - q_{b} ^{d} n_{a} n^{c} e_{cI} \omega_{d} ^{I} (n^{a} \tilde{\varepsilon}^{b} - n^{b} \tilde{\varepsilon}^{a}) + n_{a} n^{c} n_{b} n^{d} e_{cI} \omega_{d} ^{I} (n^{a} \tilde{\varepsilon}^{b} - n^{b} \tilde{\varepsilon}^{a}) \right]\d t.
\end{eqnarray}
Note that most of the terms vanishes due to $q_{a} ^{c} n^{a} = 0$ or by antisymmetry of the indices, the non vanishing terms are,
\begin{equation}
e_{aI} \omega_{b} ^{I} \tilde{\varepsilon}^{ab}  = - N \left[  q_{a} ^{c} n_{b} n^{d}  n^{b} \tilde{\varepsilon}^{a}  - q^{d} _{b} n_{a} n^{c} n^{a} \tilde{\varepsilon}^{b} \right] e_{cI} \omega_{d} ^{I}\d t.
\end{equation}
Since $n^{a}$ is the normal to the spacelike surfaces $M$ (and the splitting in the boundary is compatible with the spacetime one), $n_{a} n^{a} = -1$. Also we use $n^{a} = (t^{a} - N^{a})/N$, $\mathcal{E}_{a} ^{I} = q_{a} ^{b} e_{b} ^{I}$ and $\mathcal{W}_{a} ^{I} = q_{a} ^{b} \omega_{b} ^{I}$, the integrand of the boundary term becomes,

\begin{eqnarray}
\nonumber e_{aI} \omega_{b} ^{I} \tilde{\varepsilon}^{ab}  &=& - N \left[ \mathcal{E}_{aI} \frac{1}{N} (t^{d} - N^{d}) (n_{b} n^{b}) \omega_{d} ^{I} \tilde{\varepsilon}^{a} - \frac{1}{N} (t^{c} - N^{c} ) \omega_{d} ^{I} e_{cI} (n_{a} n^{a}) \tilde{\varepsilon}^{b}\right]\d t \\
&=& -  (n_{b} n^{b} ) \left[ t^{d} \omega_{d} ^{I} \mathcal{E}_{aI} \tilde{\varepsilon}^{a} - N^{d} \omega_{d} ^{I} \mathcal{E}_{aI} \tilde{\varepsilon}^{a} + t^{c} e_{cI} \mathcal{W}_{d} ^{I} \tilde{\varepsilon}^{d} - N^{c} e_{cI} \mathcal{W}_{d} ^{I} \tilde{\varepsilon}^{d} \right]\d t
\end{eqnarray}
which implies that the decomposed standard boundary term is,
\begin{equation}\label{ABV-boundarytermdecomposed}
 -  \frac{1}{ \kappa} \int_{\partial \mathcal{M} } e_{aI} \omega_{b} ^{I} \tilde{\varepsilon}^{ab}  = -  \frac{1}{ \kappa} \int_{\partial \mathcal{M} }  \left[ t^{d} \omega_{d} ^{I} \mathcal{E}_{aI} \tilde{\varepsilon}^{a} - N^{d} \omega_{d} ^{I} \mathcal{E}_{aI} \tilde{\varepsilon}^{a} + t^{c} e_{cI} \mathcal{W}_{d} ^{I} \tilde{\varepsilon}^{d} - N^{c} e_{cI} \mathcal{W}_{d} ^{I} \tilde{\varepsilon}^{d} \right]\d t
\end{equation}

Now we decompose the integrand of the additional boundary term (\ref{AdditionalHarmlessBoundaryTerm}), $\frac{1}{ n \cdot n} \varepsilon^{IKL} e_{aI} n_{K} \mathring{\bar{\mathcal{D}}} _{b} n_{L} \tilde{\varepsilon}^{ab}$,
\begin{eqnarray}
\nonumber \frac{1}{ n \cdot n} \varepsilon^{IKL} e_{aI} n_{K} \mathring{\bar{\mathcal{D}}} _{b} n_{L} \tilde{\varepsilon}^{ab} &=& \frac{1}{ n \cdot n} \varepsilon^{IKL} e_{cI} \delta_{a} ^{c} n_{K}  \delta_{b} ^{d} \mathring{\bar{\mathcal{D}}} _{d} n_{L} \tilde{\varepsilon}^{ab}\\
\nonumber &=&  \frac{1}{ n \cdot n} \varepsilon^{IKL} e_{cI}  n_{K} \mathring{\bar{\mathcal{D}}} _{d} n_{L} (q_{a} ^{c} - n_{a} n^{c} ) ( q_{b} ^{d} - n_{b} n^{d} ) \tilde{\varepsilon}^{ab}\\
\nonumber &=& -  \frac{1}{\sqrt{ n \cdot n}} \varepsilon^{IL}  \left[ \left( \mathcal{E}_{aI}  t^{d} \mathring{\bar{\mathcal{D}}} _{d} n_{L}   -  \mathcal{E}_{aI}  N^{d} \mathring{\bar{\mathcal{D}}} _{d} n_{L}  \right) \tilde{\varepsilon}^{a} \right. \\
&& \left. +  \left( t^{c} e_{cI}  \mathring{\bar{\mathcal{D}}} _{b} n_{L}      - N^{c} \mathcal{E}_{cI}  \mathring{\bar{\mathcal{D}}} _{b} n_{L}   \right)  \tilde{\varepsilon}^{b} \right]  \d t
\end{eqnarray}
for the previous equation we used $n^{c} = \frac{1}{N}(t^{c} - N^{c})$, $n^{a}$ is normal to a spacelike surface so $n_{a} n^{a} = -1$,  $\mathring{\bar{\mathcal{D}}} _{d}$ is spatial so $q_{b} ^{d} \mathring{\bar{\mathcal{D}}} _{d}  = \mathring{\bar{\mathcal{D}}} _{d} $, and $\mathcal{E}_{a} ^{I} = q_{a} ^{b} e_{b} ^{I}$. Thus the decomposed boundary term ((\ref{AdditionalHarmlessBoundaryTerm})) is,
\begin{eqnarray}\label{ABV-decomposition-additionalBT}
\nonumber - \frac{\alpha}{\kappa} \int_{\partial \mathcal{M} } \frac{1}{ n \cdot n} \varepsilon^{IKL} e_{aI} n_{K} \mathring{\bar{\mathcal{D}}} _{b} n_{L} \tilde{\varepsilon}^{ab} &=&  \frac{\alpha}{\kappa} \int_{\partial \mathcal{M} }  \frac{1}{ \sqrt{n \cdot n}} \varepsilon^{IL}  \left[ \left( \mathcal{E}_{aI}  t^{d} \mathring{\bar{\mathcal{D}}} _{d} n_{L}   -  \mathcal{E}_{aI}  N^{d} \mathring{\bar{\mathcal{D}}} _{d} n_{L}  \right) \tilde{\varepsilon}^{a} \right. \\
&&  + \left.  \left( t^{c} e_{cI}  \mathring{\bar{\mathcal{D}}} _{b} n_{L}      - N^{c} \mathcal{E}_{cI}  \mathring{\bar{\mathcal{D}}} _{b} n_{L}   \right)  \tilde{\varepsilon}^{b} \right]  \d t
\end{eqnarray}
Using (\ref{ABV-Bulktermdecomposed}), (\ref{ABV-boundarytermdecomposed}), (\ref{ABV-decomposition-additionalBT}) and $e = \sqrt{-g} = N \sqrt{|q|} = N \mathcal{E}$ with $q$ the determinant of the induced metric $q_{ab}$ on $M$ and $\mathcal{E}$ the determinant of $\mathcal{E} ^{a} _{I}$, we can rewrite the action (\ref{Palatini-arbitraryG}) as,
 \begin{eqnarray}
\nonumber S_{PB} [e, \omega ] &=& - \frac{1}{2 \kappa} \int \d t \int_{M} N \mathcal{E} \varepsilon^{LKI} \left[ \mathcal{E}_{L} ^{b} \mathcal{E}_{K} ^{c} \mathcal{F}_{bcI} - \frac{2}{N} \mathcal{E}_{L} ^{b} n_{K} t^{c} F_{bcI} + \frac{2}{N} \mathcal{E}_{L} ^{b} n_{K} N^{c} \mathcal{F}_{bcI} \right]\\
\nonumber &&-  \frac{1}{ \kappa} \int \d t \int_{\partial M }   \left[ t^{d} \omega_{d} ^{I} \mathcal{E}_{aI} \varepsilon^{a} - N^{d} \omega_{d} ^{I} \mathcal{E}_{aI} \varepsilon^{a} + t^{c} e_{cI} \mathcal{W}_{d} ^{I} \varepsilon^{d} - N^{c} e_{cI} \mathcal{W}_{d} ^{I} \varepsilon^{d} \right] \\
\nonumber  && +  \frac{\alpha}{\kappa} \int  \d t \int_{\partial M}  \frac{1}{ n \cdot n} \varepsilon^{IL}  \left[ \left( \mathcal{E}_{aI}  t^{d} \mathring{\bar{\mathcal{D}}} _{d} n_{L}   -  \mathcal{E}_{aI}  N^{d} \mathring{\bar{\mathcal{D}}} _{d} n_{L}  \right) \tilde{\varepsilon}^{a} \right. \\
 && \left. +  \left( t^{c} e_{cI}  \mathring{\bar{\mathcal{D}}} _{b} n_{L}      - N^{c} \mathcal{E}_{cI}  \mathring{\bar{\mathcal{D}}} _{b} n_{L}   \right)  \tilde{\varepsilon}^{b} \right] 
 \end{eqnarray}
As in the Witten decomposition, we use (\ref{tF-igual-LieD-menos-CovD}) to rewrite the second term of the bulk part of the action,
\begin{eqnarray}
\nonumber N \mathcal{E} \varepsilon^{LKI} (\frac{2}{N} \mathcal{E}_{L} ^{b} n_{K}  t^{c} F_{cb} ^{I}) \nonumber &=&  \mathcal{E} \varepsilon^{LKI} 2 \mathcal{E}_{L} ^{b} n_{K} \pounds_{\vec{t}} \omega_{b} ^{I} -  \mathcal{E} \varepsilon^{LKI}  2 \mathcal{E}_{L} ^{b} n_{K}\mathcal{D}_{b} (t\cdot \omega)^{I} \\
\nonumber &=& 2 \mathcal{E} \varepsilon^{LKI} \left[ \mathcal{E}_{L} ^{b} n_{K} \pounds_{\vec{t}} \omega_{b} ^{I} + \mathcal{D}_{b} \left( \mathcal{E}_{L} ^{b} n_{K} \right) (t\cdot \omega)^{I} \right]\\
 && - \mathcal{D}_{b} \left[ \mathcal{E} \varepsilon^{LKI}  2 \mathcal{E}_{L} ^{b} n_{K} (t\cdot \omega)^{I} \right]
\end{eqnarray}
Then the action can be written,
 \begin{eqnarray}
\nonumber S_{PB} [e, \omega ] &=& - \frac{1}{2 \kappa} \int \d t \int_{M} \left[ N \mathcal{E} \varepsilon^{LKI}  \mathcal{E}_{L} ^{b} \mathcal{E}_{K} ^{c} \mathcal{F}_{bcI} + 2 \mathcal{E} \varepsilon^{LKI} \left( \mathcal{E}_{L} ^{b} n_{K} \pounds_{\vec{t}} \omega_{b} ^{I} + \mathcal{D}_{b} \left( \mathcal{E}_{L} ^{b} n_{K} \right) (t\cdot \omega)^{I}\right. \right. \\
\nonumber && \left. \left.  +  \mathcal{E}_{L} ^{b} n_{K} N^{c} \mathcal{F}_{bcI} \right)  \right] + \frac{1}{2 \kappa} \int \d t \int_{M} \mathcal{D}_{b} \left[ \mathcal{E} \varepsilon^{LKI}  2 \mathcal{E}_{L} ^{b} n_{K} (t\cdot \omega)^{I} \right] \\
\nonumber &&- \frac{1}{ \kappa} \int \d t \int_{\partial M }    \left[ t^{d} \omega_{d} ^{I} \mathcal{E}_{aI} \varepsilon^{a} - N^{d} \omega_{d} ^{I} \mathcal{E}_{aI} \varepsilon^{a} + t^{c} e_{cI} \mathcal{W}_{d} ^{I} \varepsilon^{d} - N^{c} e_{cI} \mathcal{W}_{d} ^{I} \varepsilon^{d} \right] \\
\nonumber  && + \frac{\alpha}{\kappa} \int  \d t \int_{\partial M}  \frac{1}{ n \cdot n} \varepsilon^{IL}  \left[ \left( \mathcal{E}_{aI}  t^{d} \mathring{\bar{\mathcal{D}}} _{d} n_{L}   -  \mathcal{E}_{aI}  N^{d} \mathring{\bar{\mathcal{D}}} _{d} n_{L}  \right) \tilde{\varepsilon}^{a} \right. \\
 && \left.  +  \left( t^{c} e_{cI}  \mathring{\bar{\mathcal{D}}} _{b} n_{L}      - N^{c} \mathcal{E}_{cI}  \mathring{\bar{\mathcal{D}}} _{b} n_{L}   \right)  \tilde{\varepsilon}^{b} \right] 
 \end{eqnarray}
To find the Hamiltonian we need to calculate the momenta to perform the Legendre transformation,
\begin{equation}
\Pi_{I} ^{b} = \frac{\delta \mathcal{L} }{\delta (\pounds_{\vec{t}} \omega_{b} ^{I}  )  } = \frac{1}{\kappa}  \mathcal{E} \varepsilon^{LKI}  \mathcal{E}_{L} ^{b} n_{K} 
\end{equation}
Then,
\begin{eqnarray}\label{Hamiltonian-ABV}
\nonumber H[e, \omega] &=& \int_{M} \left[ (\pounds_{\vec{t}} \omega_{c} ^{I}  )\Pi_{I} ^{c}  - \mathcal{L} \right]\\
\nonumber &=& + \frac{1}{2 \kappa}  \int_{M} \left[ N \mathcal{E} \varepsilon^{LKI}  \mathcal{E}_{L} ^{b} \mathcal{E}_{K} ^{c} \mathcal{F}_{bcI} + 2 \mathcal{E} \varepsilon^{LKI} \left[\mathcal{D}_{b} \left( \mathcal{E}_{L} ^{b} n_{K} \right) (t\cdot \omega)^{I}  +  \mathcal{E}_{L} ^{b} n_{K} N^{c} \mathcal{F}_{bcI} \right]  \right]\\
\nonumber && - \frac{1}{2 \kappa}  \int_{M} \mathcal{D}_{b} \left[ \mathcal{E} \varepsilon^{LKI}  2 \mathcal{E}_{L} ^{b} n_{K} (t\cdot \omega)^{I} \right] \\
\nonumber &&+  \frac{1}{ \kappa} \int_{\partial M }   \left[ t^{d} \omega_{d} ^{I} \mathcal{E}_{aI} \varepsilon^{a} - N^{d} \omega_{d} ^{I} \mathcal{E}_{aI} \varepsilon^{a} + t^{c} e_{cI} \mathcal{W}_{d} ^{I} \varepsilon^{d} - N^{c} e_{cI} \mathcal{W}_{d} ^{I} \varepsilon^{d} \right]\\
 \nonumber && - \frac{\alpha}{\kappa} \int_{\partial M}  \frac{1}{ n \cdot n} \varepsilon^{IL}  \left[ \left( \mathcal{E}_{aI}  t^{d} \mathring{\bar{\mathcal{D}}} _{d} n_{L}   -  \mathcal{E}_{aI}  N^{d} \mathring{\bar{\mathcal{D}}} _{d} n_{L}  \right) \tilde{\varepsilon}^{a} \right. \\
 && \left. +  \left( t^{c} e_{cI}  \mathring{\bar{\mathcal{D}}} _{b} n_{L}      - N^{c} \mathcal{E}_{cI}  \mathring{\bar{\mathcal{D}}} _{b} n_{L}   \right)  \tilde{\varepsilon}^{b} \right] 
\end{eqnarray}
Note that within this decomposition we have `more structure', now we have three constraints
\begin{equation}
\varepsilon^{LKI}  \mathcal{E}_{L} ^{b} \mathcal{E}_{K} ^{c} \mathcal{F}_{bcI} \approx 0 ,\,\,\,\,\,  \varepsilon^{LKI} \mathcal{D}_{b} \left( \mathcal{E}_{L} ^{b} n_{K} \right) \approx 0 \,\,\,\,\,  \mathrm{and} \,\,\,\,\,  \mathcal{E}_{L} ^{b} n_{K} \mathcal{F}_{bcI} \approx 0,
\end{equation}
instead of the two found by the Witten approach (\ref{Witten-constraints}).

On the constraint surface we are left only with the boundary term,
\begin{eqnarray}
\nonumber H &=& - \frac{1}{2 \kappa}  \int_{M} \mathcal{D}_{b} \left[ \mathcal{E} \varepsilon^{LKI}  2 \mathcal{E}_{L} ^{b} n_{K} (t\cdot \omega)^{I} \right]\\
\nonumber && +  \frac{1}{ \kappa} \int_{\partial M }    \left[ t^{d} \omega_{d} ^{I} \mathcal{E}_{aI} \varepsilon^{a} - N^{d} \omega_{d} ^{I} \mathcal{E}_{aI} \varepsilon^{a} + t^{c} e_{cI} \mathcal{W}_{d} ^{I} \varepsilon^{d} - N^{c} e_{cI} \mathcal{W}_{d} ^{I} \varepsilon^{d} \right]\\
\nonumber &&- \frac{\alpha}{\kappa} \int_{\partial M}  \frac{1}{ n \cdot n} \varepsilon^{IL}  \left[ \left( \mathcal{E}_{aI}  t^{d} \mathring{\bar{\mathcal{D}}} _{d} n_{L}   -  \mathcal{E}_{aI}  N^{d} \mathring{\bar{\mathcal{D}}} _{d} n_{L}  \right) \tilde{\varepsilon}^{a} \right. \\
&& \left.  +  \left( t^{c} e_{cI}  \mathring{\bar{\mathcal{D}}} _{b} n_{L}      - N^{c} \mathcal{E}_{cI}  \mathring{\bar{\mathcal{D}}} _{b} n_{L}   \right)  \tilde{\varepsilon}^{b} \right] 
\end{eqnarray}

Let us now consider the asymptotically flat boundary conditions. The leading term of $(t \cdot \omega)^{I} = 0$ and also since $\mathring{\bar{\mathcal{D}}} _{d}$ is spatial $t^{d} \mathring{\bar{\mathcal{D}}} _{d} n_{L} = 0$ . So we are left with
\begin{eqnarray}\label{ABV-hamiltonian-boundary-threeterms}
\nonumber H &=& \lim_{r \to \infty} \left\{- \frac{1}{ \kappa} \int_{\partial M }   \left[  N^{\bar{d}} \,^{1} \mathcal{W}_{\bar{d} } ^{I} \,^{0}\mathcal{E}_{ \bar{a}I} \varepsilon^{\bar{a} } - t^{c} \,^{0}e_{cI} \,^{1} \mathcal{W}_{\bar{d} } ^{I} \varepsilon^{\bar{d} } + N^{\bar{c} } \,^{0}\mathcal{E}_{ \bar{c}I} \,^{1}\mathcal{W}_{\bar{d} } ^{I} \varepsilon^{\bar{d} } \right] \right.\\
\nonumber &&- \left. \frac{\alpha}{\kappa} \int_{\partial M}  \frac{1}{ n \cdot n} \varepsilon^{IL}  \left[  -  \,^{0}\mathcal{E}_{aI}  N^{d} \mathring{\bar{\mathcal{D}}} _{d} n_{L} \tilde{\varepsilon}^{a}  +  \left( t^{c} \,^{0}e_{cI}  \mathring{\bar{\mathcal{D}}} _{b} n_{L}      - N^{c} \,^{0}\mathcal{E}_{cI}  \mathring{\bar{\mathcal{D}}} _{b} n_{L}   \right)  \tilde{\varepsilon}^{b} \right] \right. \\
&& \left. +  {\cal O}(r^{-1}) \right\}
\end{eqnarray}
In addition to the fall-off conditions on $e$ and $\omega$, now we have to take into account the behaviour of the lapse $N$ and shift $N^{a}$ functions on the asymptotic region for time-translations (following \cite{Ashtekar-Varadarajan1994, Marolf-Patino2006}),
\begin{eqnarray}
\label{Falloff-Laps} N &=& 1 + {\cal O}(r^{-1})\\
\label{Falloff-Shift} N^{a} &=& {\cal O}(r^{-1-\beta}),
\end{eqnarray}
Note that in the asymptotic region the projections $\mathcal{E}_{a} ^{I} = q_{a} ^{b} e_{b} ^{I}$ and $\mathcal{W}_{a} ^{I} = q_{a} ^{b} \omega_{b} ^{I}$ coincide with $e_{\bar{a}} ^{I}$ and $\omega_{\bar{a}} ^{I}$. With conditions (\ref{Falloff-Laps}),(\ref{Falloff-Shift}) and considering the order of leading terms of $e$ and $\omega$: $\,^{1} \omega_{\bar{d} } ^{I} = {\cal O}(r^{-1}) = \,^{1} \mathcal{W}_{\bar{d} } ^{I}$, $\,^{0} e_{ \bar{c}I} = {\cal O}(r^{-\beta/2}) = \,^{0} \mathcal{E}_{ \bar{a}I}$, and that $\varepsilon^{\bar{d} } =  {\cal O}(r)$. Note that to first order the first and third terms in (\ref{ABV-hamiltonian-boundary-threeterms}) decay as,
\begin{eqnarray}
\lim_{r \rightarrow \infty } \frac{1}{2 \kappa} \int_{\partial M }   N^{\bar{d}} \,^{1}\omega_{\bar{d} } ^{I} \,^{0} \mathcal{E}_{ \bar{a}I} \varepsilon^{\bar{a} } &=& \lim_{r \rightarrow \infty } \frac{1}{2 \kappa} \int_{\partial M }    {\cal O}(r^{-1-\beta}) {\cal O}(r^{-1}) {\cal O}(r^{-\beta/2}) {\cal O}(r)\\
&=& \lim_{r \rightarrow \infty } \frac{1}{2 \kappa} \int_{\partial M } {\cal O}(r^{-1-3 \beta / 2}) = 0
\end{eqnarray}
and
\begin{eqnarray}
\lim_{r \rightarrow \infty } \frac{1}{2 \kappa} \int_{\partial M }   N^{\bar{c} } \,^{0}\mathcal{E}_{ \bar{c}I} \,^{1}\mathcal{W}_{\bar{d} } ^{I} \varepsilon^{\bar{d} } &=& \lim_{r \rightarrow \infty } \frac{1}{2 \kappa} \int_{\partial M }    {\cal O}(r^{-1-\beta}) {\cal O}(r^{-\beta/2}){\cal O}(r^{-1})  {\cal O}(r)\\
&=& \lim_{r \rightarrow \infty } \frac{1}{2 \kappa} \int_{\partial M } {\cal O}(r^{-1-3 \beta / 2}) = 0,
\end{eqnarray}
respectively. And the fourth and sixth terms decay as,
\begin{eqnarray}
\nonumber \lim_{r \rightarrow \infty }  \frac{\alpha}{\kappa} \int_{\partial M}  \frac{1}{ n \cdot n} \varepsilon^{IL} \left[    \,^{0}\mathcal{E}_{aI}  N^{d} \mathring{\bar{\mathcal{D}}} _{d} n_{L} \tilde{\varepsilon}^{a} \right]   &=& \lim_{r \rightarrow \infty } \frac{1}{2 \kappa} \int_{\partial M } {\cal O}(r^{-\beta/2})    {\cal O}(r^{-1-\beta}) {\cal O}(r^{-1-\beta /2})  {\cal O}(r)\\
&=& \lim_{r \rightarrow \infty } \frac{1}{2 \kappa} \int_{\partial M } {\cal O}(r^{-1-2 \beta }) = 0
\end{eqnarray}
and
\begin{eqnarray}
\nonumber \lim_{r \rightarrow \infty }  \frac{\alpha}{\kappa} \int_{\partial M}  \frac{1}{ n \cdot n} \varepsilon^{IL} \left[   N^{c} \,^{0}\mathcal{E}_{cI}  \mathring{\bar{\mathcal{D}}} _{b} n_{L}   \tilde{\varepsilon}^{b} \right]   &=& \lim_{r \rightarrow \infty } \frac{1}{2 \kappa} \int_{\partial M }    {\cal O}(r^{-1-\beta}) {\cal O}(r^{-\beta/2})  {\cal O}(r^{-1-\beta /2})  {\cal O}(r)\\
&=& \lim_{r \rightarrow \infty } \frac{1}{2 \kappa} \int_{\partial M } {\cal O}(r^{-1-2 \beta }) = 0.
\end{eqnarray}
Therefore, $H$ can be written as,
\begin{equation}
\nonumber H = \lim_{r \to \infty} \left\{- \frac{1}{ \kappa} \int_{\partial M }   \left[   - t^{c} \,^{0}e_{cI} \,^{1} \mathcal{W}_{\bar{d} } ^{I} \varepsilon^{\bar{d} } \right] - \frac{\alpha}{\kappa} \int_{\partial M}  \frac{1}{ n \cdot n} \varepsilon^{IL}  \left[ t^{c} \,^{0}e_{cI}  \mathring{\bar{\mathcal{D}}} _{b} n_{L} \tilde{\varepsilon}^{b} \right] +  {\cal O}(r^{-1}) \right\}
\end{equation}

As in the previous sections, if we want this Hamiltonian to generate asymptotic time translations and therefore its conserved quantity to be the \emph{energy}, $t^{a}$ has to approach a time-translation Killing vector field of the asymptotic flat spacetime, which also translates in $t$ being orthogonal to $M$ (corresponding to $N\to 1, N^a\to 0$). In that case 
the previous expression coincides with (\ref{H-nonvanishingleadingtermscomefrom}) from

Therefore, $H$ can be written as,
\begin{eqnarray}
\nonumber H &=& \lim_{r \to \infty} \left\{- \frac{1}{ \kappa} \int_{\partial M }   \left[   - t^{c} \,^{0}e_{cI} \,^{1} \mathcal{W}_{\bar{d} } ^{I} \varepsilon^{\bar{d} } \right] - \frac{\alpha}{\kappa} \int_{\partial M}  \frac{1}{ n \cdot n} \varepsilon^{IL}  \left[ t^{c} \,^{0}e_{cI}  \mathring{\bar{\mathcal{D}}} _{b} n_{L} \tilde{\varepsilon}^{b} \right] +  {\cal O}(r^{-1}) \right\} \\
&=& \lim_{r \rightarrow \infty} \left\{  \frac{1}{ \kappa} \int_{\partial M }   \,^{0} e_{0I} \frac{\,^{1}\bar{\omega}_{\bar{d} } ^{I} }{r} \varepsilon^{\bar{d} }  - \frac{\alpha}{ \kappa}  \int_{C_{t}}  \frac{1}{\sqrt{ n \cdot n}} \varepsilon^{IL}  \, ^{0}  e_{0I}  \left( \, ^{0} e_{cL} \mathring{\bar{\mathcal{D}}} _{b} r^{c} \right) \tilde{\varepsilon}^{b} \right\}.
\end{eqnarray}
Which is exactly the same term as (\ref{H-Wittendecomposition}), the one found by the Witten's decomposition. Therefore the Hamiltonian is the same as (\ref{Witten-lastHamiltonian}) of the previous part,
\begin{equation}
H = - \frac{ 1}{2 \kappa} \int_{C_{t}} (2 \alpha - \beta) \d \theta=\frac{1}{8G}\,(\beta - 2)\, ,
\end{equation}
which is the same result we obtained for the Witten decomposition.
Note that at the end of the day, the result for the energy is the same in both decompositions as expected, 
this is due to the fact that at the asymptotic region the direction of $t^{a}$ coincides with $n^{a}$, and 
also the lapse y shift functions decay in such a way. This may not be true for other conserved quantities 
such as the angular momentum, but we shall leave the discussion to forthcoming works.

\section{Discussion}
\label{sec:6}

In this work we have addressed the issue of defining well posed variational principles for first order asymptotically flat 2+1 gravity, {\emph{and}} their corresponding Hamiltonian descriptions, in both the covariant and canonical formalisms. Of particular relevance was the issue of recovering the Hamiltonian and therefore the energy as
a boundary term after performing the Legendre transform, {\emph{without}} the need to postulate extra
boundary terms to render the formalism consistent (as is the case in the Regge-Teitelboim formalism \cite{regge-teitel,Ashtekar-Varadarajan1994}). 
As we have shown, this question can be answered in the affirmative
not for one, but for {\emph{two}} different actions, each of which yields a different value for the energy of the spacetime. In turn, this clarifies a tension that existed in the literature regarding, say, the energy of Minkowski spacetime. One should also note that this program has only been recently completed in first order
3+1 gravity as well \cite{CR}.

Let us now summarize our results. First, we proposed a three dimensional manifestly Lorentz invariant  Palatini action $S_{\textrm{LIP}}$ that is well posed under asymptotically flat boundary conditions. As we have noted, the analogue of the well posed Palatini action in $4D$ \cite{aes}, that we called $S_{\textrm{SPB}}$, is not manifestly Lorentz invariant, although it has a well posed action principle under the asymptotically flat boundary conditions. This is so given that one has to make a partial gauge fixing in the boundary to make it invariant under the residual gauge transformations. As we showed in detail, by introducing an additional appropriate boundary term \ref{AdditionalHarmlessBoundaryTerm}, we can indeed define an action that is manifestly Lorentz invariant and moreover, this action coincide with the three dimensional Einstein-Hilbert action with a Gibbons-Hawking term. We derived the asymptotically flat boundary conditions for the first order variables, and with these conditions we showed that the proposed action $S_{\textrm{LIP}}$ has a well posed variational principle, i.e., it is finite and differentiable. Then, using the covariant and canonical approaches we obtained an expression for the energy. In the first case, the covariant formalism  can at best yield an expression for the {\it{variation}} of the energy. Thus, our results are analogous to those in \cite{Ashtekar-Varadarajan1994} where the Regge-Teitelboim method was used for the second order metric variables. In the second case, using a canonical formalism, we could directly compare our results with those in \cite{Marolf-Patino2006}, where the starting point is the Einstein-Hilbert action with Gibbons-Hawking term, that is well posed under asymptotically flat boundary conditions.

To summarize, we have two results: When we start with the action $S_{\textrm{SPB}}$, the 
corresponding boundary contribution yields a {\it positive} energy in the interval 
$[0,1/4G]$. Thus, Minkowski spacetime is assigned zero energy. When we consider
the manifestly gauge invariant action $S_{\textrm{LIP}}$ obtained by the
addition of the term \ref{AdditionalHarmlessBoundaryTerm}, we recover the results 
of \cite{Marolf-Patino2006}. Namely, in this case the gravitational energy is always negative and contained within the interval $[-1/4G,0]$. Thus, Minkowski spacetime has a negative energy equal to $-1/4G$.

Let us now end with some remarks regarding these results.

\begin{enumerate}
\item As is also standard practice in asymptotically flat 3+1 gravity, we have focused our attention on the gravitational action, without considering any particular matter content. This does not mean that our considerations are restricted to the vacuum case. The assumption that we have made, as is done in the 3+1 case, is that the decay rates of matter fields are stronger in such a way that there is no contribution to the boundary terms of the action coming from the matter fields. Thus, the Hamiltonian does not depend explicitly on the matter fields.\footnote{Recall that the situation is similar in 3+1 gravity. Even when the expression for energy depends explicitly only on geometrical fields, these depend through Einstein's equations on the matter content.
Even more, the vacuum 2+1 case would be trivial.} Thus, the expressions for energy we have found are valid for generic matter content (satisfying reasonable energy conditions).

\item Let us compare our results here regarding the different actions with the situation in 3+1 gravity. In 3+1, the standard second order action for asymptotically flat spacetimes is given by the Einstein-Hilbert bulk term of the form $\int_M R$ plus a boundary term of the form $\int_{\partial M}(K-K_0)$, where one subtracts a non-dynamical (infinite) term to make the action finite (See, however \cite{mm} for a discussion of the viability of this action).  In the first order formalism, the
Palatini action plus a simple boundary term \cite{aes}, the analogue to our $S_{\textrm{SPB}}$ action, is already finite and has been shown to be related, under certain conditions to the finite second order action \cite{aes}.
In 2+1 gravity, the action of the form
$\int_M R + \int_{\partial M}K$ is already finite and does not need to be `renormalized', as shown in \cite{Marolf-Patino2006}. Here we have shown that the
totally gauge invariant action $S_{\textrm{LIP}}$ is equal to the Marolf-Pati\~no action $\int_M R + \int_{\partial M}K$. Moreover, just as in the 3+1 case, the action
$S_{\textrm{SPB}}$ that we considered here is a `shifted' version of the Marolf-Patiño action. The difference with the 3+1 case is that, in 2+1 dimensions, this non-dynamical `shift' is finite rendering both actions well defined, while in the 3+1 case only one of them is viable.

\item In 3+1 gravity, several arguments strongly suggest that the ADM four momentum of Minkowski spacetime should vanish. On the one hand, there is no combination of the fundamental constants of the theory (for simple matter content) that has dimensions of mass, so it would be unnatural to have a non-zero value for energy of the vacuum configuration. Even more, symmetry considerations suggest that a Poincare invariant configuration (in terms of asymptotic symmetries) should have zero ADM four-momentum. Otherwise, a non-zero ADM four-vector would select a preferred (asymptotic) frame, violating Poincare invariance. In three dimensions, none of this features exist. To begin with, the gravitational constant $G$ has dimensions of inverse mass. Second, since the asymptotic metric is not that of Minkowski spacetime but that of a cone (flat with a deficit angle), translations are {\it not} a symmetry of the asymptotic spacetime \cite{deser2,symmetry}. Thus, a preferred frame is not in principle excluded. Given all this, it is not surprising nor completely unexpected that Minkowski spacetime might have a non-zero value for energy.

\item As we have mentioned, the asymptotic symmetry group of AF 2+1 gravity is qualitatively different 
from the 3+1 case. Two distinct lines of research have been pursued to study the structure of this group.
In \cite{symmetry}, conformal techniques were employed to describe such symmetries. In \cite{barnich} a 
different strategy, motivated by work on AdS was put forward. It would be interesting to take our 
Hamiltonian description as a staring point, and systematically study the structure of the asymptotic 
symmetries. This will be left for a future publication. 

\item As was early noted \cite{Witten1988}, at the level of actions for a compact spatial slice, the 
Einstein-Palatini action is equivalent to a Chern-Simons theory for the group $ISO(2,1)$. They differ, 
precisely, by a boundary term. The natural question is whether one can define a consistent action by 
adding appropriate boundary terms to the bulk Chern-Simons form. Furthermore, one would like to study the 
same issues we have considered here, and obtain the energy as defined by that action. 
This shall be reported elsewhere \cite{CR-CS}. 

\end{enumerate}

\appendix



\section{On the new boundary term}\label{NewBoundaryTerm-constant}

As we commented on previous sections, particularly in section \ref{sec:3}, the addition of the term (\ref{AdditionalHarmlessBoundaryTerm}),
\begin{equation}
\int_{\partial \mathcal{M} } \frac{1}{ n \cdot n} \varepsilon^{IKL} e_{I} \wedge n_{K} \mathrm{d} n_{L}
\end{equation}
has many advantages. It is necessary for the action to be manifestly Lorentz invariant and it has a constant value when evaluated on histories compatible with the asymptotically flat boundary conditions, so it does not spoil finiteness nor differentiability. The resulting well posed manifestly Lorentz invariant action is equivalent to the Einstein Hilbert action so we can fully recover previous results obtained by means of the metric formulation. In the appendices we shall prove this assertions.

Here $n_{K}$  is a spacetime scalar that is an internal vector. We can define it by $n_{K}\, / \sqrt{n \cdot n} := R^{a} e_{aK}$ where $R^{a}$ is the spacetime unit normal to the boundary\footnote{Note that we have extended the usual definition of $n_{K} = n^{a} e_{aK}$ for the Cauchy surfaces in the first order formalism to $n_{k}\, / \sqrt{n \cdot n} := R^{a} e_{aK}$ that allows, in principle, $n_{K}$ to be rescaled, and now is extended also to include the timelike boundary.}, that can either be $n^{a}$ for the unit normal to the spacelike surfaces or $r^{a}$ for the unit normal to the timeline boundary, we have introduced a normalization factor $\frac{1}{n \cdot n}$ to allow freedom in rescaling $n _{K}$, so we can use any multiple of $n_{K}$ and the results will remain the same. Since $n_{K}$ is a spacetime scalar $\mathrm{d} n_{L}$ is a one form as well as $e_{I}$ then the previous boundary term is the integral of a two form over a two dimensional boundary.

For the more general case, when the boundary might become null one needs to use densitized internal normals as discussed in \cite{bn}, such that the expressions do not diverge. In the case treated here it is enough and more intuitive to use just the $n_{K}$.

\subsection{New boundary term evaluated on Asymptotically flat boundary conditions}

In this subsection we shall prove that the term (\ref{AdditionalHarmlessBoundaryTerm}) is constant when evaluated on the boundary conditions. On the boundary, the term (\ref{AdditionalHarmlessBoundaryTerm}) can be written as,
\begin{eqnarray}
\int_{\partial \mathcal{M} } \frac{1}{ n \cdot n} \varepsilon^{IKL} e_{I} \wedge n_{K} \mathrm{d} n_{L}  &=& \int_{\partial \mathcal{M} } \frac{1}{ n \cdot n} \varepsilon^{IKL} e_{aI} n_{K} \mathring{\bar{\mathcal{D}}}_{b} n_{L} \tilde{\varepsilon}^{ab}\\
&=&  \left[ - \int_{M_{1}} + \int_{M_{2}} + \int_{\mathcal{I}} \right] \frac{1}{ n \cdot n} \varepsilon^{IKL} e_{aI} n_{K} \mathring{\bar{\mathcal{D}}}_{b} n_{L} \tilde{\varepsilon}^{ab}
\end{eqnarray}

Where we are  considering  the region $\mathcal{M}$  bounded by $\partial_{\mathcal{M}} = M_{1} \cup M_{2} \cup \mathcal{I}$, $M_{1}$ and $M_{2}$ are space-like slices and $\mathcal{I}$ an outer boundary.  Recall that we choose the torsion free flat connection $\mathring{\bar{\mathcal{D}}}_{b}$, such that $D = \mathring{\bar{\mathcal{D}}} + \omega$ and $\mathring{\bar{\mathcal{D}}}_{b} \, ^{0} \bar{e}_{a} ^{I} = 0$ and also that $n_{k}\, / \sqrt{n \cdot n} := R^{a} e_{aK}$ where $R^{a}$ is the spacetime unit normal to the boundary, that can either be $n^{a}$ for the unit normal to the spacelike surfaces or $r^{a}$ for the unit normal to the timelike boundary. For the timelike part,
\begin{eqnarray}
\int_{ \mathcal{I} } \frac{1}{ n \cdot n} \varepsilon^{IKL} n_{K} e_{I} \wedge  \mathrm{d} n_{L}  &=& \int_{ \mathcal{I} }  \underbrace{(\varepsilon^{IKL}\frac{n_{K}}{\sqrt{n \cdot n }} )}_{-\varepsilon^{IL}} e_{aI}  \mathring{\bar{\mathcal{D}}}_{b} \underbrace{ \left( \frac{n_{L}}{\sqrt{ n \cdot n}} \right)}_{r^{a} e_{aK}} \tilde{\varepsilon}^{ab}\\
&=& - \int_{\mathcal{I} }  \varepsilon^{IL} e_{aI} \left( r^{c} \mathring{\bar{\mathcal{D}}}_{b}  e_{cL} + e_{cL}  \mathring{\bar{\mathcal{D}}}_{b} r^{c}  \right) \tilde{\varepsilon}^{ab}\\
&=& \underbrace{- \int_{ \mathcal{I} }  \varepsilon^{IL} e_{aI}  r^{c} \mathring{\bar{\mathcal{D}}}_{b}  e_{cL} }_{B_{1}} \underbrace{ - \int_{\mathcal{I} } \varepsilon^{IL} e_{aI}  e_{cL}  \mathring{\bar{\mathcal{D}}}_{b} r^{c}  \tilde{\varepsilon}^{ab}}_{B_{2}}.
\end{eqnarray}

From the previous equation we have two terms, $B_{1}$ and $B_{2}$. We shall analyze first $B_{1}$, when evaluated on the boundary the term becomes,
\begin{eqnarray}\label{B1-evaluationofSuperTerm}
B_{1} &=& \lim_{r \rightarrow \infty} \left[ - \frac{\alpha}{ \kappa}  \int_{ \mathcal{I}}   \varepsilon^{IL} \, ^{0} e_{aI}  r^{c} ( - \frac{\beta}{2r} r^{ - \beta /2 } \partial_{b} r \, ^{0} \bar{e}_{\bar{c}L} \delta^{\bar{c}} _{c} )\tilde{\varepsilon}^{ab} + {\cal O}(r^{- 1})\right] \\
\nonumber &=& \lim_{r \rightarrow \infty}\left[   \frac{\alpha \beta}{ 2\kappa}  \int_{ \mathcal{I}}  \underbrace{\varepsilon^{IL} \, ^{0} \bar{e}_{aI} \, ^{0} \bar{e}_{\bar{c}L}}_{\bar{e} \tilde{\varepsilon}_{a\bar{c} }}   \frac{1}{r} r^{ - \beta /2 } r^{\bar{c}}  \partial_{b} r    \tilde{\varepsilon}^{ab}  + {\cal O}(r^{- 1})\right] \\
\nonumber &=& \lim_{r \rightarrow \infty} \left[   \frac{\alpha \beta}{ 2\kappa}  \int_{ \mathcal{I}}     \frac{1}{r} r^{ - \beta /2 } r^{\bar{c}}  \partial_{b} r \underbrace{ \tilde{\varepsilon}_{a\bar{c} } \tilde{\varepsilon}^{ab} }_{\delta_{\bar{c}} ^{b}}   r \d\theta \d t + {\cal O}(r^{- 1})\right] \\
\nonumber &=& \lim_{r \rightarrow \infty} \left[   \frac{\alpha \beta}{ 2\kappa}  \int_{\mathcal{I}}      r^{ - \beta /2 } (+1)   \d\theta \d t + {\cal O}(r^{- 1})\right] \\
\nonumber &=& \lim_{r \rightarrow \infty} \left[   \frac{\alpha \beta}{ 2\kappa} r^{ - \beta /2 } 2\pi + {\cal O}(r^{- 1})\right]  \\
\nonumber &=& \lim_{r \rightarrow \infty}  \left[  {\cal O}(r^{- \beta / 2})  + {\cal O}(r^{- 1})\right]  = 0   \,\,\,\,\, iff\,\,\,\,\, \beta > 0\\
\end{eqnarray}
and $B_{2}$ becomes,
\begin{eqnarray}\label{B2-evaluationofSuperTerm}
B_{2} &=&  \lim_{r \rightarrow \infty}  \left[ - \frac{\alpha}{ \kappa}  \int_{\mathcal{I}}  \varepsilon^{IL}  \, ^{0}  e_{0I}  \left( \, ^{0} e_{cL} \mathring{\bar{\mathcal{D}}} _{b} r^{c} \right) \tilde{\varepsilon}^{b} \right]\\
\nonumber &=& - \frac{\alpha}{ \kappa} \lim_{r \rightarrow \infty}  \int_{\mathcal{I}}   \underbrace{ \varepsilon^{IL}  \, ^{0}  e_{0I}  \, ^{0} e_{cL} }_{\bar{e}  \tilde{\varepsilon}_{0c} } \underbrace{\mathring{\bar{\mathcal{D}}} _{b} r^{c}}_{\partial_{b} r^{c}}  \tilde{\varepsilon}^{0b} \\
\nonumber &=& - \frac{\alpha}{ \kappa} \lim_{r \rightarrow \infty}  \int_{\mathcal{I}}  \underbrace{ \tilde{\varepsilon}^{0b} \tilde{\varepsilon}_{0c}}_{\delta^{b} _{c}} (\partial_{b} r^{c} ) r \d\theta \d t\\
&=& - \frac{\alpha}{ \kappa} \lim_{r \rightarrow \infty}  \int_{\mathcal{I}} \underbrace{ (\partial_{c} r^{c} )}_{1/r} r \d \theta  \d t=  - \frac{\alpha}{2 \kappa}  \int_{\mathcal{I}} 2 \d \theta  \d t
\end{eqnarray}

Therefore the value of the boundary term (\ref{AdditionalHarmlessBoundaryTerm})  when evaluated in the timelike boundary and on the boundary conditions becomes,
\begin{eqnarray}
\int_{\mathcal{I} } \frac{1}{ n \cdot n} \varepsilon^{IKL} n_{K} e_{I} \wedge  \mathrm{d} n_{L}  
&=&B_{1} + B_{2}\\
&=& \lim_{r \rightarrow \infty}  \left[  {\cal O}(r^{- \beta / 2})  + {\cal O}(r^{- 1})\right]  - \frac{\alpha}{2 \kappa}  \int_{\mathcal{I}} 2 \d \theta \\
&=& - \frac{\alpha}{2 \kappa}  \int_{\mathcal{I}} 2 \d \theta  \d t.
\end{eqnarray}
Since we are integrating over a finite time interval with $M_{1}$ and $M_{2}$ asymptotically time-translated with respect to each other, the previous integral take a finite constant value.

Analogously, we can follow the same steps but for $R^{a} = n^{a}$ and check that the boundary term corresponding to the spacelike surfaces is also constant. Thus, the whole boundary term is constant when evaluated on the boundary conditions.


\section{On the equivalence between second order and first order actions}
 \label{Appendix-Equiv-Einstein-vs-Palatini}

It has been shown for the three dimensional Einstein-Hilbert action that the Gibbons-Hawking term is the only term needed to make the variational principle well posed \cite{Marolf-Patino2006}. Taking $\kappa = 8 \pi G$, the Einstein-Hilbert action with Gibbons-Hawking term is,
\begin{equation}
S_{\textrm{EH-GH}} [g] = \frac{1}{2 \kappa} \int_{\mathcal{M} } \sqrt{-g} R + 2 \int_{\partial \mathcal{M} } \sqrt{-h} K
\end{equation}
with $R$ the Ricci scalar, $g$ the determinant of the spacetime metric $g_{ab}$, $h$ the determinant of the induced metric on the boundary $\partial \mathcal{M}$ and $K$ the extrinsic curvature of the boundary.

We shall prove, on the other hand, that the Lorentz invariant well posed Palatini action with boundary term,
\begin{equation}
S_{\textrm{LIP}}[e,\omega] = - \frac{1}{ \kappa} \int_{\mathcal{M}} e^{I} \wedge F_{I} \; \; - \frac{1}{ \kappa} \int_{\partial \mathcal{M} } \frac{1}{n \cdot n} \varepsilon^{IKL} e_{I} \wedge n_{K} \mathcal{D} n_{L}.
\end{equation}
is in fact equivalent to the Einstein-Hilbert action with Gibbons-Hawking term.

We study first the Einstein-Hilbert term, $\frac{1}{2 \kappa} \int_{\mathcal{M} } \sqrt{-g} R$, considering that $g^{ab} = \eta^{IJ} e^{a} _{I} e^{b} _{J} $, $\sqrt{-g} = e $, $2e e^{[a|} _{I} e^{|c]} _{J}  = \tilde{\eta}^{acf} \varepsilon_{IJK}e^{K} _{f}  $, $F_{ab} ^{IJ} =  e^{cI} e^{dJ} R_{acbd} $ and $F_{ab} ^{JK} = F_{ab} ^{L} \varepsilon ^{KJ} \, _{L}$. The bulk term,
\begin{eqnarray}
\nonumber \frac{1}{2 \kappa} \int_{\mathcal{M} } \sqrt{-g} R &=&  \frac{1}{2 \kappa} \int_{\mathcal{M} } \underbrace{\sqrt{-g} }_{e} \underbrace{g^{ab}}_{ \eta^{IJ} e^{a} _{I} e^{b} _{J} } \underbrace{R_{ab}}_{ R_{acbd} g^{cd} } \\
\nonumber &=& \frac{1}{2 \kappa} \int_{\mathcal{M} } e e^{[a|} _{I } e^{bI} R_{acbd} e^{|c]} _{J } e^{dJ}\\
\nonumber &=& \frac{1}{2 \kappa} \int_{\mathcal{M} } \frac{1}{2} \underbrace{2 e e^{[a|} _{I } e^{|c]} _{J }}_{\tilde{\eta}^{acf} \varepsilon_{IJK}e^{K} _{f}}  e^{bI} e^{dJ}  R_{acbd} \\
\nonumber &=& \frac{1}{2 \kappa} \int_{\mathcal{M} } \frac{1}{2} \tilde{\varepsilon}^{acf} \varepsilon_{IJK}e^{K} _{f}
\underbrace{e^{bI}  e^{dJ} R_{acbd}}_{F_{ac} ^{IJ} } \\
\nonumber &=& \frac{1}{2 \kappa} \int_{\mathcal{M} } \frac{1}{2} \tilde{\varepsilon}^{acf} \varepsilon_{IJK}e^{I} _{f}
\underbrace{F_{ac} ^{JK}}_{F_{ac} ^{L} \varepsilon ^{KJ} \, _{L}}  \\
\nonumber &=& \frac{1}{2 \kappa} \int_{\mathcal{M} } \frac{1}{2} \tilde{\varepsilon}^{acf} \underbrace{\varepsilon_{IJK} \varepsilon ^{KJ} \, _{L} }_{- 2 \delta_{I} ^{L} }  e^{I} _{f}
F_{ac} ^{L}   \\
&=& - \frac{1}{2 \kappa} \int_{\mathcal{M} } \tilde{\varepsilon}^{acf} e_{f} ^{I} F_{acI}    \\
&=& - \frac{1}{ \kappa} \int_{\mathcal{M}} e^{I} \wedge F_{I}
\end{eqnarray}
Note the change in sign when we write down the Palatini action  defined over an arbitrary Lie group (see e.g. \cite{romano}).

Now we shall see the relation between the Lorentz invariant boundary term (\ref{AppLI-Result}) introduced 
in section \ref{sec:3} and the Gibbons Hawking term. We begin with the Lorentz invariant boundary term,
\begin{equation}
 \int_{\partial \mathcal{M} } \frac{1}{n \cdot n} \varepsilon^{IKL} e_{I} \wedge n_{K} \mathcal{D} n_{L} =  \left[ - \int_{M_{1}} + \int_{M_{2}} + \int_{\mathcal{I}} \right] \frac{1}{n \cdot n} \varepsilon^{IKL} e_{I} \wedge n_{K} \mathcal{D} n_{L}
\end{equation}
where our integration region $\mathcal{M}$ is bounded by $\partial_{\mathcal{M}} = M_{1} \cup M_{2} \cup \mathcal{I}$, $M_{1}$ and $M_{2}$ are space-like slices and $\mathcal{I}$ a family of timelike cylinders we used to approach spatial infinity.

For the timelike boundary consider $n_{L} / \sqrt{n \cdot n}:= r^{a} e_{aL}$, $r^{a}$ the normal to the cylinder, $\mathcal{D}_{c} r^{a} = \nabla_{c} r^{a}$ where $\nabla$ is the Levy Civita connection, $\gamma_{ab}$ is the induced metric on the timelike boundary and that $\varepsilon^{IKL} e_{bI}  e_{dK}   e_{aL} = e \tilde{\varepsilon}_{bda}$. The term on the timelike boundary is,
\begin{eqnarray}
\nonumber  \int_{\mathcal{I} } \frac{1}{n \cdot n} \varepsilon^{IKL} e_{I} \wedge n_{K} \mathcal{D} n_{L} &=&  \int_{\mathcal{I} }  \varepsilon^{IKL} e_{bI}  \frac{n_{K}}{\sqrt{n \cdot n}} \mathcal{D}_{c} \left( \frac{n_{L}}{\sqrt{n \cdot n}} \right) \tilde{\varepsilon}^{bc}\\
\nonumber &=&  \int_{\mathcal{I} }  \varepsilon^{IKL} e_{bI} \frac{ n_{K}}{\sqrt{n \cdot n}} \mathcal{D}_{c} (r^{a} e_{aL}) \tilde{\varepsilon}^{bc}\\
\nonumber &=&  \int_{\mathcal{I} }  \varepsilon^{IKL} e_{bI}  \frac{n_{K}}{\sqrt{n \cdot n}} \left[ r^{a} \underbrace{\mathcal{D}_{c}  e_{aL}}_{=0 \,\,by\,\, EOM} +  e_{aL} \mathcal{D}_{c} r^{a}  \right] \tilde{\varepsilon}^{bc}\\
\nonumber &=&  \int_{\mathcal{I} }  \varepsilon^{IKL} e_{bI} (r^{d} e_{dK})   e_{aL} \mathcal{D}_{c} r^{a}  \tilde{\varepsilon}^{bc}\\
\nonumber &=& \int_{\mathcal{I} }  \varepsilon^{IKL} e_{bI}  e_{dK}   e_{aL}  r^{d} \nabla_{c} r^{a}  \tilde{\varepsilon}^{bc}\\
\nonumber &=& - \int_{\mathcal{I} } e \underbrace{(\tilde{\varepsilon}_{bda}  r^{d})}_{- \varepsilon_{ab}} \nabla_{c} r^{a} \varepsilon^{bc} \sqrt{-\gamma}\\
\nonumber &=&-  \int_{\mathcal{I} } \sqrt{-\gamma}  \nabla_{c} r^{a} \underbrace{(- \tilde{\varepsilon}_{ab}  \tilde{\varepsilon}^{bc} )}_{\delta_{a} ^{c}}\\
&=& -  \int_{\mathcal{I} } \sqrt{-\gamma}  \nabla_{a} r^{a}. 
\end{eqnarray}
 Now we can recall that we define the extrinsic curvature, $\mathcal{K}$,  of a surface (in this case the timelike cylinder) as the trace of $\mathcal{K}_{a} ^{b} = \nabla_{a} r^{b}$ where $r^{b}$ is the normal to the surface, then $\mathcal{K} = \gamma^{ab} \mathcal{K}_{ab} = \mathcal{K}_{a} ^{a} = \nabla_{a} r^{a}$. With this at hand we can see 
that, in fact,  
\begin{equation}
\int_{\mathcal{I} } \frac{1}{n \cdot n} \varepsilon^{IKL} e_{I} \wedge n_{K} \mathcal{D} n_{L}  = - \int_{\mathcal{I} } \sqrt{-\gamma}  \nabla_{a} r^{a} = -  \int_{\mathcal{I} } \sqrt{-\gamma}  \mathcal{K},
\end{equation}
where $\mathcal{K}$ is the extrinsic curvature of the timelike boundary. Following an analogous derivation for the spacelike surfaces $M_{1}$ and $M_{2}$, we can easily see that,
\begin{equation}
 \int_{M_{1,2} } \frac{1}{n \cdot n} \varepsilon^{IKL} e_{I} \wedge n_{K} \mathcal{D} n_{L} = -  \int_{M_{1,2} } \sqrt{q}  \nabla_{a} n^{a} = -  \int_{M_{1,2} } \sqrt{q} \, k,
\end{equation}
again, with $q$ the determinant of the induced metric on $M_{1,2}$, $n^{a}$ and $k$ its normal vector and extrinsic curvature respectively. With this at hand we can see that,
\begin{equation}
 \int_{\partial \mathcal{M} } \frac{1}{n \cdot n} \varepsilon^{IKL} e_{I} \wedge n_{K} \mathcal{D} n_{L} = -  \left[ - \int_{M_{1}} + \int_{M_{2}}\right] \sqrt{q} \, k  - \int_{\mathcal{I}} \sqrt{-\gamma}  \mathcal{K}  = -  \int_{\partial \mathcal{M} } \sqrt{-h} K.
\end{equation}

From (\ref{AppLI-Result}) in the section \ref{sec:3}, we can see that,
\begin{eqnarray}
 \frac{1}{ \kappa}  \int_{\partial \mathcal{M} } \sqrt{-h} K 
&=& - \frac{1}{\kappa}  \int_{\partial \mathcal{M} } \frac{1}{n \cdot n} \varepsilon^{IKL} e_{I} \wedge n_{K} \mathcal{D} n_{L}\\
\label{AppEHGWvsPB-1} &=& - \frac{1}{\kappa} \int_{\partial \mathcal{M}} e^{I} \wedge \omega_{I} -  \frac{1}{\kappa} \int_{\partial \mathcal{M} } \frac{1}{ n \cdot n} \varepsilon^{IKL} e_{I} \wedge n_{K} \mathrm{d} n_{L}
\end{eqnarray}
This result coincides, apart from the second term of the right hand side of the last equation, with that given in \cite{Miskovic2006} when the cosmological constant is zero. In \cite{Miskovic2006} they use the Gaussian (normal) coordinates and also they consider particular internal directions for the spin connection. This ``fixing" of the internal directions is reflected in the fact that the second term of the RHS in (\ref{AppEHGWvsPB-1}) is not present in their action.

\section*{Acknowledgments}
\noindent
This work was in part supported by DGAPA-UNAM IN103610 grant, by CONACyT 0177840 
and 0232902 grants, by the PASPA-DGAPA program, by NSF
PHY-1403943 and PHY-1205388 grants, and by the Eberly Research Funds of Penn State.


\end{document}